\documentclass[%
aps,
 reprint,
groupedaddress,
showpacs,preprintnumbers,
amsmath,amssymb,
longbibliography,
pra,
]{revtex4-1}

\usepackage[utf8]{inputenc}
\usepackage{graphicx}
\usepackage{dcolumn}
\usepackage{bm}
\usepackage{bbold}
\usepackage{braket}
\usepackage{tikz,pgfplots}
\usepackage{tikz-qtree}
\usepackage{xcolor}
\usepackage{units}
\usepackage{tabularx}
\usepackage{booktabs}
\usepackage[outdir=./]{epstopdf}
\pgfplotsset{compat=1.10}

\usetikzlibrary{patterns,shadows,fadings,positioning,trees,calc}
\usetikzlibrary{shapes}

\tikzfading[name=fade cloud,
         inner color=transparent!99,
         outer color=transparent!40]
\tikzfading[name=fade inside,
         inner color=transparent!80,
         outer color=transparent!30]
\tikzfading[name=fade out,
         inner color=transparent!0,
         outer color=transparent!90]
\tikzfading[name=fade inner,
         inner color=transparent!100,
         outer color=transparent!0]

\DeclareMathOperator\erf{erf}

\definecolor{diplom1}{RGB}{101 156 239}
\definecolor{diplom2}{RGB}{000 000 128}
\definecolor{diplom3}{RGB}{153,0,0} 
\definecolor{diplom4}{RGB}{232,215,23}
\definecolor{diplom5}{RGB}{51,37,60}

\definecolor{unirot}{RGB}{153,0,0}
\definecolor{unirot_hell}{RGB}{255,228,225}
\definecolor{lightblue}{RGB}{242.2,249.88,255}

\AtBeginDocument{
\heavyrulewidth=.08em
\lightrulewidth=.05em
\cmidrulewidth=.03em
\belowrulesep=.65ex
\belowbottomsep=0pt
\aboverulesep=.4ex
\abovetopsep=0pt
\cmidrulesep=\doublerulesep
\cmidrulekern=.5em
\defaultaddspace=.5em
}

\begin{document}


\title{Time-resolved Spectroscopy of Interparticle Coulombic Decay Processes}

\author{Elke Fasshauer}
 \email{elke.fasshauer@gmail.com}
\author{Lars Bojer Madsen}%
\affiliation{%
 Department of Physics and Astronomy, Aarhus University\\
 Ny Munkegade 120, 8000 Aarhus, Denmark
}%

\date{\today}

\begin{abstract}
We report theory for time-resolved spectator resonant Interparticle Coulombic
Decay (ICD) processes. Following excitation by a short extreme ultraviolet pulse,
the spectrum of the resonant ICD electron develops. Strong-field ionization is
imagined to quench the decay at different time delays and to initiate
regular ICD.
In this latter process, the ICD electron signal can be measured without interference effects. The typical lifetimes of ICD processes allow for the observation of oscillations of the time- and energy-differential ionization probability. We propose to utilize this oscillation to measure lifetimes of electronic decay processes.
\end{abstract}

\maketitle

\section{Introduction}
Interparticle Coulombic Decay (ICD) \cite{Cederbaum97,Marburger03}
is an electronic decay process of ionized systems
consisting of at least two weakly bonded units, be it atoms, molecules or clusters.
ICD has been observed
in multiple systems like noble gas clusters \cite{Santra00_1,Hergenhahn11,
Jahnke15,Fasshauer16, Fasshauer10, Fasshauer13, Fasshauer14_1, Foerstel16, Fasshauer17}
and clusters of different solvent molecules like
water \cite{Mueller06,Hergenhahn11,Jahnke15}
or ammonia \cite{Kryzhevoi11_1,Stoychev11,Kryzhevoi11_2,Oostenrijk18}.
It allowed to explain the repair mechanism of the enzyme
photolyase \cite{Harbach13} and was used to establish a more efficient double ionization
strategy \cite{Stumpf16a}. ICD is furthermore discussed as
a source of slow electrons, which are most efficient
in damaging the DNA after exposure to high energy radiation or radioactive materials in the
human body \cite{Boudaiffa00, Brun09, Pan03, Martin04, Surdutovich12, Alizadeh15}.
In this work, we shed light on this fundamental process by offering a time-resolved
perspective directly of the electron dynamics.

\begin{figure*}[t]
 \begin{center}
 \includegraphics[width=17cm]{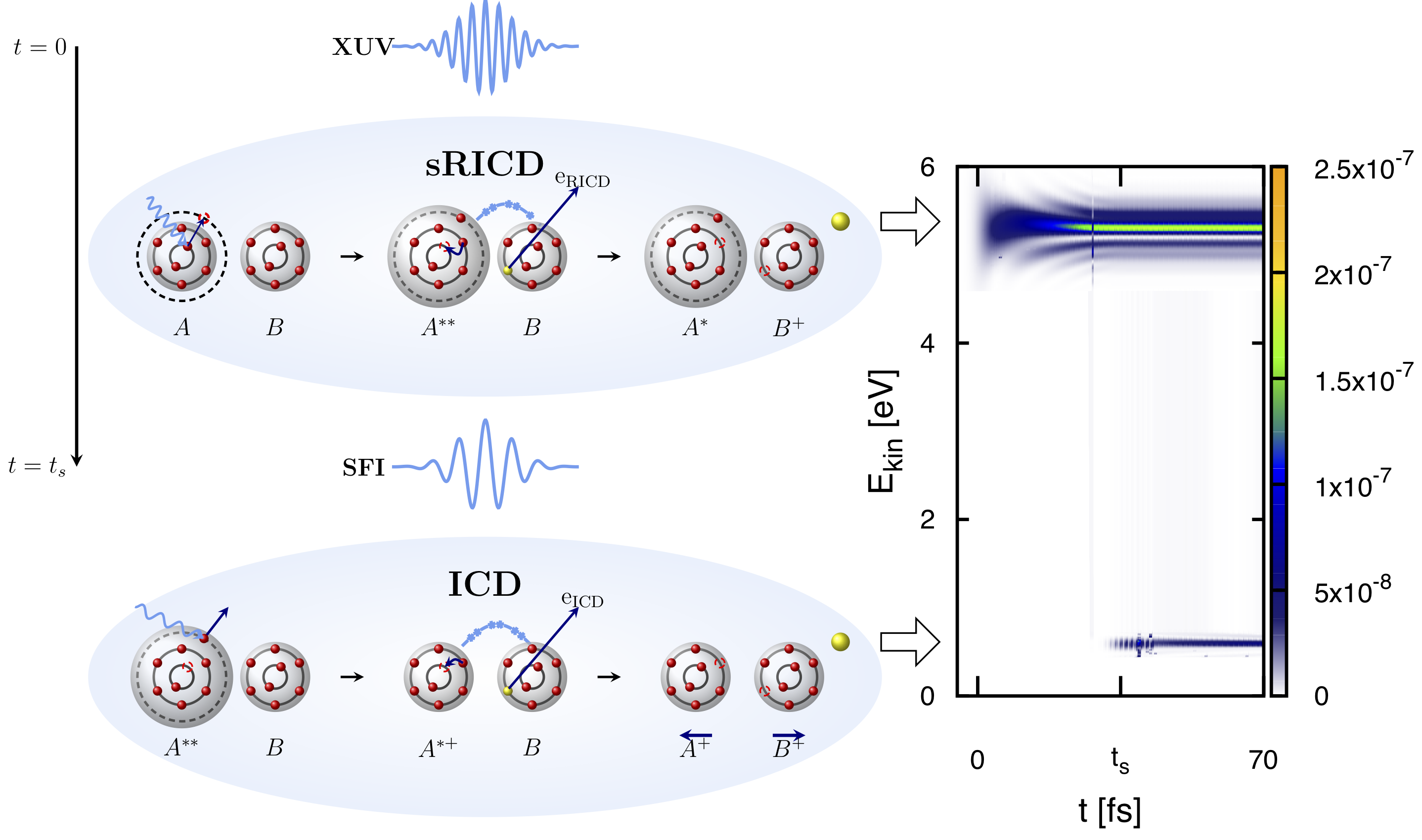}
 \end{center}
 \caption{Illustration \cite{rapid_spec} of the considered process. 
 At $t=0$, a spectator Resonant
          Interparticle Coulombic Decay (sRICD) is initiated
          by XUV excitation. At a later time $t_s$, which in this example plot is
          \unit[35]{fs}, a second
          strong infrared laser pulse quenches the sRICD process by ionization and thereby
          initiates ICD. The signals of the respective electrons with
          kinetic energies
          $E_{\mathrm{kin}}$ are well separated. While the sRICD appears from the start of the
          XUV pulse centered at $t=0$, the ICD signal appears after the second pulse.
          This approach allows for a
          time-resolved measurement of the sRICD signal and an interference free measurement
          of the ICD signal.
          \textbf{sRICD}:
          Unit $A$ is excited from the inner valence shell.
          The created vacancy is filled
         with an electron from the outer valence and the excess energy is simultaneously
         transferred to $B$, which emits the sRICD electron.
         \textbf{ICD}: The initial state is created by removing the excited electron
         in $A$. It is then filled by an electron of the outer valence of $A$ while the excess energy is transferred to $B$, which
         consequently emits the ICD electron. The two units are both positively charged
         and therefore undergo Coulomb explosion.
         The panel to the right illustrates typical time and energy resolved traces
         of sRICD and ICD electrons (see text).
         }
 \label{fig:rapid_spec}
\end{figure*}

Due to developments in creating short pulses in the extreme ultraviolet (XUV) and
x-ray domain \cite{Krausz09}
a time-dependent
investigation of ICD processes should be within reach.
Fano profiles of a much faster
autoionization (AI) process were recently measured \cite{Kaldun16,Gruson16}.
A few time-resolved investigations of ICD
have already been performed theoretically and experimentally,
where the ions produced in the process were measured
\cite{Kuleff07,Schnorr13,Trinter13a,Schnorr15,Fruehling15,Mizuno17,Takanashi17,Qingli19}.
However, 
electrons can be measured with higher energy resolution than ions and grant direct
access to the electron dynamics and interference effects during the process.
In this work, we will therefore focus on the evolution of the time- and
energy-differential ionization probability, propose how it
might be possible to measure this quantity in
experiment, and 
discuss
how decay lifetimes can be determined from the
time-resolved signal.

In brief, the ICD process starts from a unit $A$, which is ionized in the
inner valence shell.
This vacancy is filled by an electron
from the same unit and the excess energy is simultaneously transferred to a
neighbouring unit $B$.
The latter is consequently ionized by emission of the ICD electron.
The two positively charged units undergo a Coulomb explosion
(see Fig.~\ref{fig:rapid_spec}). A related  process is initiated by an excitation of
unit $A$. This resonant ICD (RICD) can be characterized by the behaviour of the excited
electron: it either participates in the decay process or not. The respective processes are
called participator RICD (pRICD) \cite{Gokhberg06}
and spectator RICD (sRICD) \cite{Barth05,aoto2006,Gokhberg06,Kopelke09,Knie14,Hans17}.
In this work, we will focus on the sRICD signal:
Unit $A$ is excited from the inner valence. The vacancy is then filled by an electron
from the valence and the excess energy is used to ionize the neighbouring unit $B$
(see Fig.~\ref{fig:rapid_spec}).
This process is usually characterized by lifetimes of several tens to hundreds of
femtoseconds.
After the sRICD process, the excited unit $A$ decays via fluorescense within
a few nanoseconds. AI and pRICD are competing
decay channels and their effect on the sRICD signal is taken into account in the
theory developed below.

We propose to initiate a sRICD process
by exciting with a short XUV pulse.
The system will then decay under emission of an sRICD
electron. This is illustrated in Fig.~\ref{fig:rapid_spec}
for electronic energies and resonance parameters corresponding to the neon dimer
after an excitation to the
Ne 2$s^{-1}5p$ state. At a later time $t_s$, a second
short and intense infrared (IR) laser pulse
quenches the sRICD process by ionization. By
varying the time delay $t_s$ one
should be able to observe the time-dependent formation of the sRICD signal,
as illustrated to the right in Fig.~\ref{fig:rapid_spec} before the second
laser pulse, by pump-probe
spectroscopy similar to
the AI process measured in Ref.~\cite{Kaldun16}.
At the same time, the proposed quenching would initiate an ICD process involving
the excited ion $A^{+*}$. Due to the
strong-field ionization initiating this latter process,
the signals are well-separated in energy.
In this paper we provide the basic formulation for a purely electronic solution
upon which more complex scenarios, including nuclear dynamics, will be built in future
work.

\section{Theory}
The relevant property for the description of the time evolution of the ICD processes
is the time- and energy-differential ionization probability obtained from the time-dependent wavefunction $| \Psi(t) \rangle$ by 
$P(E_\text{kin},t) = \sum\limits_i| \langle E_i | \Psi(t) \rangle | ^2.$
Here $\ket{E_i}$ denotes a continuum state with energy $E_i$, which entails both the
kinetic energy $E_\text{kin}$ of the emitted electron and the energy of the
final cationic state $i$.
These continuum states are orthogonal to all bound states of the system and amongst
each other.
Because the signals of the competing decay channels do not overlap, we can and will
focus on the sRICD signal.
Atomic units are used throughout unless stated otherwise.

We obtain the wavefunction of the system by 
solving the
time-dependent Schrödinger equation
 $i \partial_t \ket{\Psi(t)} = H(t) \ket{\Psi(t)}$
for a model system
consisting of the ground state $\ket{G}$, the resonant
state $|R\rangle$ for the sRICD and the resonant state $|I\rangle$ for the ICD process,
as well as three sets of continuum states characterized by the respective
final state energy of the decay process. All parameters describing a general resonant
state are designated by the index $r$.
The Hamiltonian consists of the single configuration Hamiltonian $H_0$,
the residual configuration interaction operators $V_{\text{RICD}}$ and $V_{\text{ICD}}$ as well
as $V_o$ for the competing autoionization and pRICD decay channels,
and the operator $H_{X}(t)$ of the exciting XUV
pulse: 
\begin{equation}
\label{eq:H}
 H(t) = H_{0} + V_{\text{sRICD}} + V_o + V_{\text{ICD}} + H_{X}(t).
\end{equation}
The effect of the ionizing infrared pulse is modelled by terminating
the sRICD and starting the ICD process at the time of
the second pulse $t_s$. We will come back to this point below.

\subsection{Description of the RICD process}

\begin{figure}[h]
 \centering
 \includegraphics[width=\columnwidth]{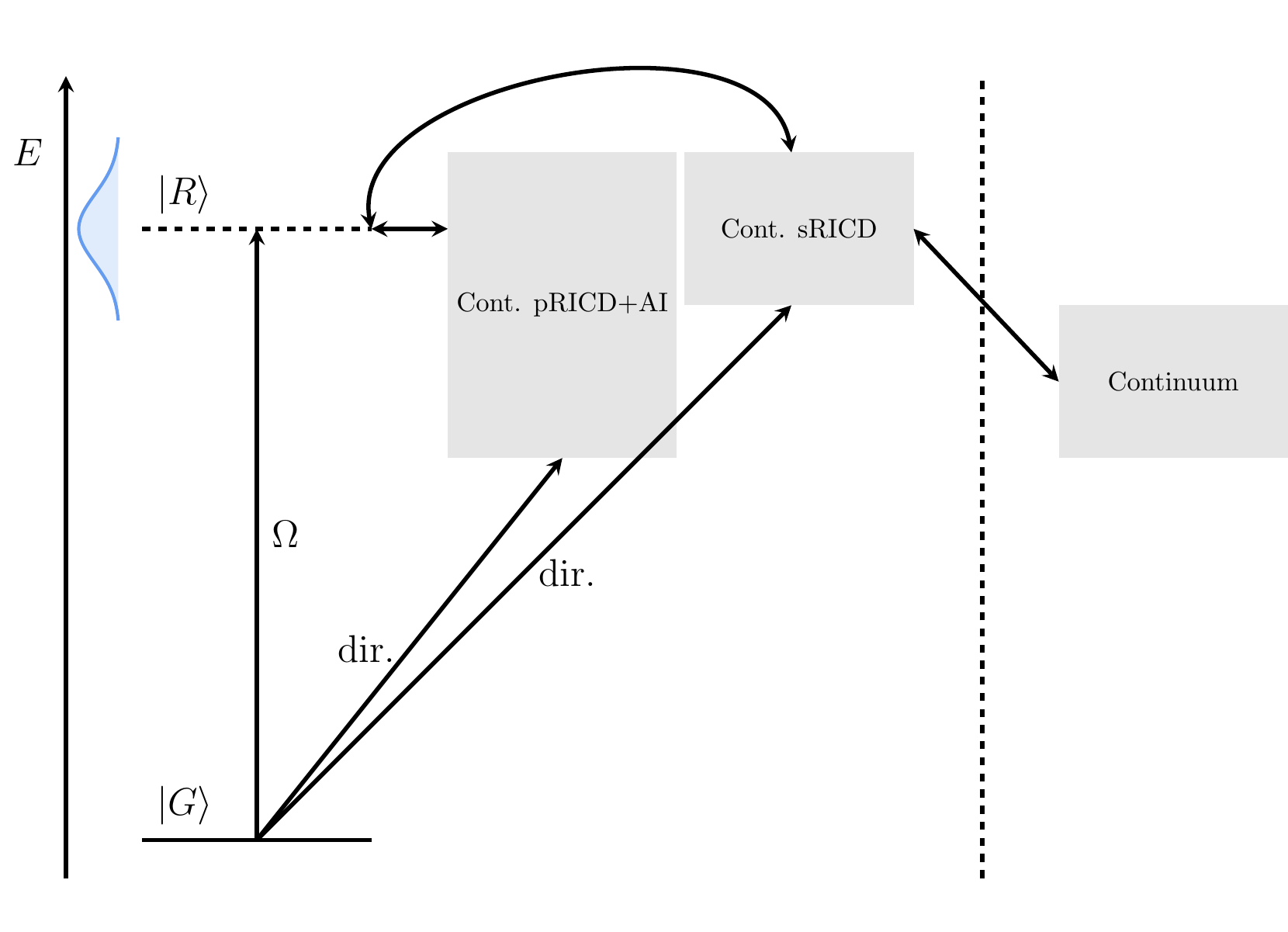}
 \caption{Schematic illustration of the energy levels involved in the RICD process.
          The laser pulse with mean photon energy $\Omega$ excites the system
          into the resonant state $\ket{R}$ with energy $E_R$.
          The resonant state couples to two continua (denoted Cont. in the figure),
          one characterized by the
          sRICD final state, the other characterized by the pRICD and AI final state.
          Both channels have partial lifetimes in the order of femtoseconds.
          The continuum state of the sRICD process
          can then couple to other continuum states
          by emission of a photon.
          This fluorescence process has a lifetime $\tau_{\text{fl.}}$,
          which typically is in the order of nanoseconds and therefore
          much larger than
          the timescales of interest here.
          Alternatively, all continuum states can be reached directly by simulateneously
          ionizing and exciting in case of the sRICD or single ionization in
          case of the pRICD and AI. All couplings to the
          continuum via spontaneous radiative transitions
          are ignored in this work.}
 \label{fig:ricd_energy_scheme}
\end{figure}

The sRICD process we intend to describe in a fully time-dependent manner
is schematically illustrated in Fig. \ref{fig:ricd_energy_scheme}.
Starting from the ground state $\ket{G}$ we excite the system with an XUV pulse with
a mean photon energy $\Omega$ into the resonant state $\ket{R}$ with the corresponding
energy $E_R$. This resonant state can then under emission of an electron
either decay to the continuum state characterized by a final state
of the sRICD $\ket{E_{sR}}$ with a partial
lifetime of $\tau_{sR}$ or to a continuum state characterized by
the shared final state of pRICD and autoionization (AI) decay $\ket{E_{o}}$  with a
different partial lifetime $\tau_o$.
The continuum state of the sRICD couples to other continuum states under emission
of a photon. This fluorescence process is usually several orders of magnitudes slower
than the electronic
decay process and is therefore ignored in this work.
Alternatively to arriving in the continuum via the decaying resonant state, a simultaneous
direct excitation and ionization in the case of sRICD
or a direct ionization in case of the pRICD and AI
channel, is also possible, which is denoted "dir" in the figure.

We assume low field strengths of the XUV pulse and therefore use first-order
perturbation theory to describe its interaction with the system. Hence, the wavefunction evolves according to

\begin{equation} \label{eq:first_order_psi}
 \ket{\Psi(t)} = \tilde{U}(t,t_0) \ket{G(t_0)}
                 -i \int\limits_{t_0}^{t} \mathrm{d}t' \, \tilde{U}(t,t') H_X(t')
                 \ket{G(t')} .
\end{equation}
Here, $\tilde{U}(t,t_0)$ is the time evolution operator from a time $t_0$ until time $t$ of the unperturbed 
system pertaining to the first four terms on the right hand side of Eq.~(\ref{eq:H}).


We are free to choose an energy reference and therefore set $E_G=0$. 
This conveniently
removes the time-dependence from the ground state, since
$\ket{G(t_0)} = \exp\{ -i E_G t_0 \} \ket{G} = \ket{G}$.

We introduce approximations for the time-evolution operator $\tilde{U}(t,t')$
appearing in the last term of Eq.~(\ref{eq:first_order_psi}).
After the system has interacted with the XUV via $H_X(t')$ at time $t'$,
the system decays via the configuration interaction $V$.
In this general discussion, $V$ refers to the sum of all configuration interaction
terms of Eq.~(\ref{eq:H}). We will specify the explicit form of $V$ for the different
cases below.
In the continuum we neglect the Coulomb interaction for simplicity,
which leads to  the time evolution operator
\begin{equation} \label{eq:Utilde}
 \tilde{U}(t, t') = U_0(t,t') - i \int\limits_{t'}^t \mathrm{d}t'' \,
                                  U_0(t,t'') \, V \, \tilde{U}(t'',t'),
\end{equation}
where  $U_0(t,t')$ is the free-particle time-evolution
operator.

By inserting Eq.(\ref{eq:Utilde}) into Eq.(\ref{eq:first_order_psi}) we arrive at
\begin{align}
\begin{split}
 \ket{\Psi(t)} =& \tilde{U}(t,t_0) \ket{G} \\
                & -i \int\limits_{t_0}^{t} \mathrm{d}t' \,
                  U_0(t,t') H_X(t') \ket{G}\\
                & -\int\limits_{t_0}^{t} \mathrm{d}t'
                   \int\limits_{t'}^t \mathrm{d}t'' \,
                  U_0(t,t'') \, V \, \tilde{U}(t'',t') \, H_X(t') \ket{G} .
\end{split}
\end{align}

The time-evolution operator $\tilde{U}(t'',t')$ in the last integral describes the
contribution of the Hamiltonian $H_0 + V$.
This is equivalent to the Hamilton
operator used in Fano's description \cite{Fano61} of a decaying resonant state
and below we therefore add the
subscript $F$ to the time-evolution operator and remove the tilde.

We now project on a continuum state $\ket{E_{sR}}$ characterized by the
sRICD final state energy
of the ionized system and the kinetic energy of the emitted electron.
The first term vanishes,
because of the orthogonality between bound and continuum states, and we obtain

\begin{align}
 \braket{E_{sR}|\Psi(t)} =& -i \int\limits_{t_0}^{t} \mathrm{d}t' \,
                        \braket{E_{sR}| U_0(t,t') H_X(t') |G} \nonumber \\
                     & -\int\limits_{t_0}^{t} \mathrm{d}t'
                        \int\limits_{t'}^t \mathrm{d}t'' \,
                        \bra{E_{sR}} U_0(t,t'') \, V \nonumber \\
                     &  \times U_F(t'',t') \, H_X(t') \ket{G} .
\label{eq:ampl_tops}
\end{align}

The time-evolution operator $U_0(t,t')$ solves the time-dependent Schrödinger equation
for the Hamiltonian $\vec{p}^2/2$. The corresponding wavefunctions with
wavevector $\vec{k}$ are
$\ket{\Psi_{\vec{k}}^0 (t)}
  = \ket{\vec{k}} \exp\biggl[ -i \int\limits_{-\infty}^t \mathrm{d}t' \frac 12 \vec{k}^2 \biggr]$.
Another way of formulating a time-evolution operator is by a projection from basis functions
at time $t'$ to such of time $t$:

\begin{equation}
 U_0(t,t') = \int \mathrm{d}\vec{k} \, \ket{\Psi_{\vec{k}}^0 (t)} \bra{\Psi_{\vec{k}}^0 (t')} .
\end{equation}

At this point we only consider the kinetic energy of the emitted electron
and can therefore
write the projection of a general continuum state $\bra{E}$ on the free-particle
time-evolution operator in the following way:

\begin{align}
 \bra{E} U_0(t,t')        =& \bra{E} \, \exp[ i\Phi_0 (E,t,t') ]\\
                          =& \bra{E} \, \exp \biggl[ -i \int\limits_{t'}^{t}
                             \left( \frac{k^2}{2} + E_\text{fin} \right)
                             \mathrm{d}t''' \biggr] , \label{eq:volkov}
\end{align}
where ${k^2}/{2}$ is the kinetic energy relative to the final state
$E_\text{fin}$.
The interaction $V$ between a general resonant state $\ket{r}$ coupling to one
of the continuum states
is given by

\begin{equation}\label{eq:V}
 V = \int \mathrm{d}E' \, \ket{E'} V_{E'r} \bra{r}
     + \int \mathrm{d}E' \, \ket{r} V_{rE'} \bra{E'} .
\end{equation}

For two different final states, and therefore two continua, it reads

\begin{align}\label{eq:V}
 V =& 
       \int \mathrm{d}E_1' \, \ket{E_1'} V_{E_1'r} \bra{r}
     + \int \mathrm{d}E_1' \, \ket{r} V_{rE_1'} \bra{E_1'} \nonumber \\
     &+ \int \mathrm{d}E_2' \, \ket{E_2'} V_{E_2'r} \bra{r}
     + \int \mathrm{d}E_2' \, \ket{r} V_{rE_2'} \bra{E_2'}.
\end{align}
This case is, e.g., relevant for
systems that can not only decay via sRICD but
also via autoionization or pRICD, which both yield a singly ionized system, which
is the case that we investigate. At the same time, the initiating energy of the XUV
laser pulse will be chosen too low to initiate an ICD process.
Hence, $V=V_{\text{sRICD}} + V_o$ in our case.

We also insert the resolution of the identity
between the Fano time-evolution $U_F(t'',t')$ operator,
to be specified below, and the Hamilton operator
of the XUV field using
\begin{align}
 \mathbb{1} &= \ket{R} \bra{R} + \int\mathrm{d}E_1' \ket{E_1'}\bra{E_1'} 
              + \int\mathrm{d}E_2' \ket{E_2'}\bra{E_2'}  \\
            &= \ket{R} \bra{R} + \int\mathrm{d}E_{sR}' \ket{E_{sR}'}\bra{E_{sR}'} 
              + \int\mathrm{d}E_o' \ket{E_o'}\bra{E_o'} \label{eq:unity}.
\end{align}

By suppressing all dependences on angular momentum and using
Eqs. (\ref{eq:ampl_tops}), (\ref{eq:V}) and (\ref{eq:unity}) we arrive at an amplitude
of the sRICD process for the resonance $| R \rangle$
consisting of four terms:

\begin{widetext}
\begin{align}
 \braket{E_{sR}|\Psi(t)} =& -i \int\limits_{t_0}^{t} \mathrm{d}t' \, \braket{E_{sR}| U_0(t,t') H_X(t') |G} 
                     -\int\limits_{t_0}^{t} \mathrm{d}t'
                        \int\limits_{t'}^t \mathrm{d}t'' \,
                       \braket{E_{sR}| U_0(t,t'') |E_{sR}} \, V_{ER} \,
                       \braket{R| U^R_F(t'',t')| R} \braket{R| H_X(t') |G} \nonumber \\
                     & -\int\limits_{t_0}^{t} \mathrm{d}t'
                        \int\limits_{t'}^t \mathrm{d}t'' \int \mathrm{d}E_{sR}' \,
                       \braket{E_{sR}| U_0(t,t'') |E_{sR}} \, V_{ER} \,
                       \braket{R| U^R_F(t'',t')| E_{sR}'} \braket{E_{sR}'| H_X(t') |G} \nonumber \\
                     & -\int\limits_{t_0}^{t} \mathrm{d}t'
                        \int\limits_{t'}^t \mathrm{d}t'' \int \mathrm{d}E_{o}' \,
                       \braket{E_{sR}| U_0(t,t'') |E_{sR}} \, V_{ER} \,
                       \braket{R| U^R_F(t'',t')| E_o'} \braket{E_o'| H_X(t') |G} .
\label{eq:ampl_three}
\end{align}
\end{widetext}

In these expressions, $U_0(t,t')$ is the free particle
time evolution operator, whose action was specified in Eq.~(\ref{eq:volkov}).
$V_{ER} = \braket{E|V|R} = \sqrt{\Gamma_{sR}/(2\pi)}$ is related to the partial
sRICD decay rate
of the resonant state $\Gamma_{sR}$ and $U^R_F(t'',t')$ is the Fano time-evolution
operator, which is specific to the resonant state $\ket{R}$.
The variables of the other, competing, processes are designated with the index $o$.
The total decay width of the RICD resonant state is given by
$\Gamma_R = \Gamma_{sR} + \Gamma_o$.

The terms of Eq.~(\ref{eq:ampl_three}) are linked to the different pathways shown
in Fig.~\ref{fig:ricd_energy_scheme} as follows.
The first term of Eq.~(\ref{eq:ampl_three}) describes the direct excitation
and ionization to the continuum state, while the second term
describes the excitation from the ground state to the resonant state followed by a decay
to the continuum state. The third, indirect, term describes the direct excitation and ionization to
the continuum state related to the sRICD final state,
which couples to the resonant state, which then again decays
into the continuum state. The last term is similar to the indirect term,
but couples to the continuum of the competing pRICD and AI channels.
Due to a different model system including only one final state,
the latter term is not present in Eq. (11) of Ref. \cite{Wickenhauser05}.
The interference of these different terms leads to the characteristic
Fano profile~\cite{Fano61}.

We describe the interaction between the system and the exciting XUV field
in the dipole approximation.
Therefore, the corresponding  Hamilton operator in the length gauge is given by
$H_X(t') = -\mu E_X(t') = -\mu \frac{\text{d}}{\text{d}t'} A_X(t') f_X(t')$,
where $E_X(t')$ denotes the time-dependent field strength of the XUV laser,
while $\mu$ denotes the dipole operator, $A_X(t') = A_{0X} \cos(\Omega t')$
is the vector potential of the laser
field in the direction of the linear polarization
and $f_X(t')$ is the Gaussian pulse envelope.
Over the energy range of interest, we assume the transition dipole matrix element from the ground state to the continuum
to be independent of the energy of the continuum state. To indicate these
assumptions, we change the notation as
$\braket{E_i|\mu|G} = \braket{C_i|\mu|G}$.
We furthermore assume the coupling elements between the resonant and the continuum state
to be real and independent of the continuum energy and therefore change the notation
$V_{E_1R} \rightarrow V_R$ and $V_{E_2R} \rightarrow W_R$.
Then, the Fano matrix elements, i.e., $\braket{R| U^R_F(t'',t')| R}$ and
$\braket{R| U^R_F(t'',t')| E'}$ of Eq.~(\ref{eq:ampl_three}),
can be solved using the projection formulation of the
Fano time-evolution operator and solving the resulting integral by contour integration.
Hence, we write the Fano time-evolution operator as

\begin{align}
 U_F(t'',t') =& \int \mathrm{d}\underline{E} \, \ket{\Psi_{\underline{E}}(t'')}
                                                \bra{\Psi_{\underline{E}}(t')} \nonumber\\
             =& \int \mathrm{d}\underline{E} \,
                \ket{\Psi_{\underline{E}}}
                \bra{\Psi_{\underline{E}}} \,
                \exp[-i\underline{E} (t''-t')] .
\end{align}

The evaluation of these Fano matrix elements is given in Appendix \ref{sec:app_fano} and
for the sRICD signal we finally arrive at:

\begin{align}
 \braket{E|\Psi(t)} =& 
                    i \braket{C_{sR}|\mu|G}
                       \int\limits_{t_0}^{t} \mathrm{d}t' \, \exp[ i\Phi_0(E_{sR},t,t') ]
                       E_X(t') \nonumber  \\
                    +& \left( V_R \braket{R|\mu|G} -i \pi V_R^2 \braket{C_{sR}|\mu|G} \right.
                       \nonumber \\
                    & \left. -i \pi V_R  W_R \braket{C_o|\mu|G} \right)  \nonumber  \\
                    &\times \int\limits_{t_0}^{t} \mathrm{d}t'
                        \int\limits_{t'}^t \mathrm{d}t'' \,
                       \exp[ i\Phi_0(E_{sR},t,t'') ] \,
                       \nonumber \\
                    &  \exp[ -i (E_R - i\pi (V_R^2 + W_R^2)) (t'' - t')] \,
                       E_X(t') . 
\label{eq:three_compact}
\end{align}

The first integral corresponds to the direct ionization process. The second integral
is identical for the resonant and the two indirect terms. They have, though, different
prefactors.
For the case of a slowly varying envelope function
($\frac{\text{d}}{\text{d}t'} f_X(t') \approx 0$) and only considering the
absorption of an XUV photon these two integrals are solved analytically for times
after the exciting pulse.
The direct term is given by
\begin{widetext}
\begin{align}
 &- \frac{A_{0X} \Omega \braket{C_{sR}|\mu|G}}{4} \,
   \exp\biggl[ -it(E_{\text{kin}} + E_{\text{fin}}) \biggr] \,
   \exp\biggl[ -\frac{\sigma^2}{2} (E_{\text{kin}} + E_{\text{fin}} -\Omega)^2 \biggr] \,
   \Re \Biggl[  \erf
   \biggl( \frac{1}{\sqrt{2} \sigma} \Bigl( \frac{T_X}{2}
   + i\sigma^2 \bigl( E_{\text{kin}} + E_{\text{fin}} - \Omega \bigr)
   \Bigr) \biggr) \Biggr]  ,
\end{align}
\end{widetext}
where $\Re$ denotes the real value and $\erf$ denotes the error function.
The resonant and indirect ionization terms without their respective prefactors
are given by
\begin{widetext}
\begin{align}
 &+ \frac{A_{0X} \, \Omega}
         {4 (E_R - i\pi (V_R^2 + W_R^2) - E_{\text{kin}} - E_{\text{fin}})} \,
   \exp\biggl[ -it(E_R - i\pi (V_r^2 + W_R^2)) \biggr] \,
   \exp\biggl[ -\frac{\sigma^2}{2} (E_R -\Omega)^2 \biggr] \,
   \Biggl( \erf(\tau_\text{1,max}) - \erf(\tau_\text{1,min}) \Biggr)             \nonumber\\
 &- \frac{A_{0X} \, \Omega}
         {4 (E_R - i\pi (V_R^2 + W_R^2) - E_{\text{kin}} - E_{\text{fin}})} \,
   \exp\biggl[ -it(E_{\text{kin}} + E_{\text{fin}}) \biggr] \,
   \exp\biggl[ -\frac{\sigma^2}{2} (E_{\text{kin}} + E_{\text{fin}} -\Omega)^2 \biggr] \,
   \Biggl( \erf(\tau_\text{2,max}) - \erf(\tau_\text{2,min}) \Biggr)  \label{eq:res_indir_expl}
\end{align}
\end{widetext}
with
$\tau_\text{1,max} = \frac{1}{\sqrt{2}\sigma}
 \Bigl( \frac{T_X}{2} - i\sigma^2(E_R -i\pi (V_R^2 + W_R^2) - \Omega) \Bigr)$,
$\tau_\text{1,min} = -\frac{1}{\sqrt{2}\sigma}
 \Bigl( \frac{T_X}{2} + i\sigma^2(E_R -i\pi (V_R^2 + W_R^2) - \Omega) \Bigr)$,
$\tau_\text{2,max} = \frac{1}{\sqrt{2}\sigma}
 \Bigl( \frac{T_X}{2} - i\sigma^2(E_{\text{kin}} + E_{\text{fin}} - \Omega) \Bigr)$ and
$\tau_\text{2,min} = -\frac{1}{\sqrt{2}\sigma}
 \Bigl( \frac{T_X}{2} + i\sigma^2(E_{\text{kin}} + E_{\text{fin}} - \Omega) \Bigr)$.
Here, $\sigma$ is the standard deviation of the Gauss distribution in time and $T_X$ is
the duration of the exciting XUV pulse.

The time- and energy-differential ionization probability of the RICD process
is the absolute square of the amplitude given in Eq. (\ref{eq:three_compact})
and therefore a sum over 28 different terms (7 absolute squares and 21 mixed 
interference terms).
Their relative contributions are determined by the Fano parameter
$q=\frac{\braket{R|\mu|G}}{ \braket{C|\mu|G} \pi V_R }$.

\subsection{Transition from the Resonant State of the RICD Process to the
            initial state of the ICD Process}
We propose a time-resolved measurement of the RICD process, where the quenching of the
RICD process by strong field ionization with an IR pulse at time $t_s$
entails the initiation of an ICD process. Technically, this means a population transfer
from the resonant state of the RICD process to the initial state of the ICD process.
We assume the transition in time to be Gaussian shaped and that the ICD signal is
unaffected by pulse shape effects.
The transition is modelled as follows.

We determine the remaining population of the resonant state of the RICD process $N_0$
at the beginning of the second pulse $t = t_s -\delta t$ as:
\begin{align}
 N_0 =& \frac{|\braket{R|\mu|G}|^2}{4} \exp[ -\sigma^2 (\Omega-E_R)^2 ] \nonumber \\
      & \times \exp[ - (\Gamma_{sR} + \Gamma_o)(t_s -\delta t) ] .
\end{align}
Here, $\delta t$ is chosen as $\frac 52 \sigma_{\text{IR}}$ with $\sigma_{\text{IR}}$
being the
standard deviation of the IR pulse in time.
We assume the decrease of the population $N_R(t)$ in the resonant state of
the RICD process $\ket{R}$
to have a Gaussian shape $f_{\text{IR}}(t)$ centered around the time of the second pulse:
\begin{equation}
 \frac{\mathrm{d}N_R(t)}{\mathrm{d}t} = -f_{\text{IR}}(t) N_R(t)
\end{equation}
and therefore
\begin{equation}
 N_R(t) = N_0 \exp \Biggl[ -\frac\alpha2 \erf \Bigl( -\frac{\delta t}{\sqrt{2}\sigma},
                                                \frac{t-t_s}{\sqrt{2}\sigma} \Bigr) \Biggr] .
\end{equation}

Correspondingly, the population of the initial state $\ket{I}$ of the ICD process
$N_R(t)$ is increased
\begin{equation}
 N_I(t) = N_0 - N_R(t) .
\end{equation}

The square roots of the time-dependent populations are then updated in every time step
and used instead of $\braket{r|\mu|G}$ in the calculation of the amplitude
of the ionization probability. In order to allow for a complete depopulation
of the resonant state, we have chosen $\alpha = 8$.

Since we assume a direct population of the resonant state of the ICD process $\ket{I}$
from the
resonant state of the sRICD process $\ket{R}$ induced
with a short and intense laser pulse,
we assume, that
the terms of Eq.~(\ref{eq:ampl_three}), which are mediated by the continuum, can be
neglected.
After the end of the ionizing pulse
the amplitude of the 
ICD process associated with resonance $| I \rangle$ therefore reads
\begin{widetext}
\begin{equation}
\label{eq:ampl_ICD}
\braket{E|\Psi(t)} = -\int\limits_{t_0}^{t} \mathrm{d}t_s
                        \int\limits_{t_s + \delta t}^t \mathrm{d}t'' \,
                       \braket{E| U_0(t,t'') |E} \, V_{EI} \,
                       \braket{I| U^I_F(t'',t_s)| I} \sqrt{N_I}.
 \end{equation}
\end{widetext}

\section{Computational Details}

\begin{table}[h]
 \caption{Energies and lifetimes for the excited states
          Ne$2s^{-1} 2p^6 np$ - Ne ($n=4,5$) undergoing RICD. The resonant energies are
          taken from Ref.~\cite{NIST2018}, the final state energies of the sRICD
          Ne$2s^2 2p^5(P_{3/2,1/2}) np$ - Ne$2p^5 (P_{3/2,1/2})$,
          of the pRICD and AI Ne$2s^2 2p^5(P_{3/2,1/2})$ - Ne
          and the ICD process
          were averaged to give
          approximations of non-relativistic energies.
          The lifetimes, $\tau$, of the sRICD and pRICD + AI
          are from Ref.~\cite{Kopelke09} and the
          ICD lifetime are from Ref.~\cite{Ghosh14b}.
         }
 \centering
 \begin{tabular}{lrrrr}
  \toprule
                                              & $n=4$     & $n=5$   & pRICD + AI        & ICD\\
  \midrule
   $E_r$ [eV]                                 & 47.1230   & \multicolumn{2}{c}{47.6930} & 48.4750\\
   $E_{\mathrm{fin}}$ [eV]                    & 41.8391   & 42.4138 & 21.6290           & 47.8688\\
   $\tau (2s^{-1} (np_z) ^1\Sigma^+_u)$ [fs]  & 112       & 106     & 206               & 98\\
  \bottomrule
 \end{tabular}
 \label{tab:param}
\end{table}

\begin{figure}[h]
 \centering
 \includegraphics[width=0.5\textwidth]{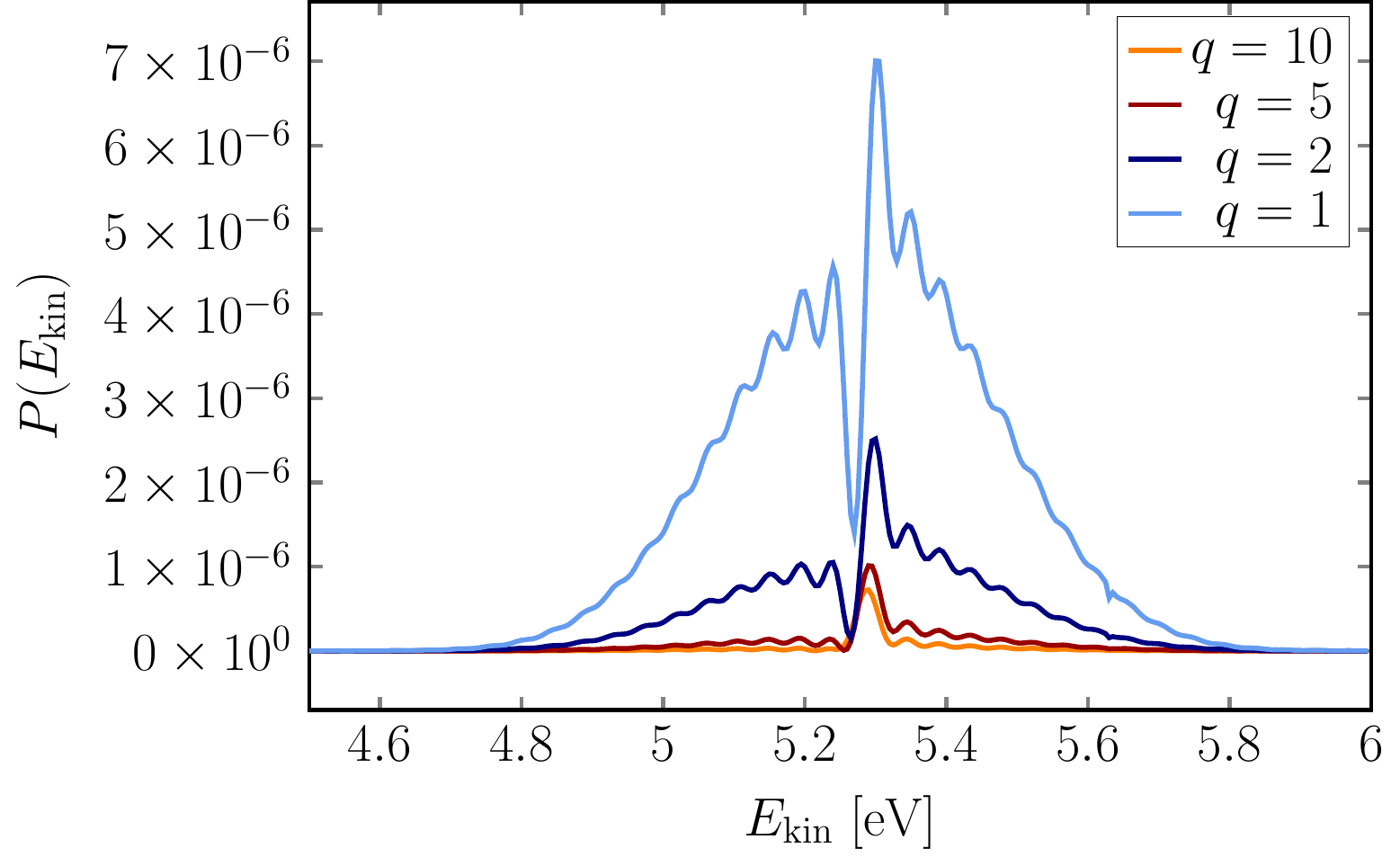}
 \caption{Peak shape of the spectator RICD electron spectrum for 50 cycles of the
          exciting pulse with different values of the Fano $q$ parameter at time
          $t=\unit[93]{fs}$ after the excitation at time $t=0$.
          The Fano profile is modified by an oscillation in energy.
          This oscillation stems from the interference terms of the direct term
          with the resonance terms
          of $P(E_{\mathrm{kin}}, t)$ and is described by Eq. (\ref{eq:oscillation}).
          The larger $q$, the less likely the direct path
          and therefore, for increasing $q$,
          the signal is damped for a given transition dipole moment
          into the resonant state.
          }
 \label{fig:nx50_q}
\end{figure}

In our model system, where the nuclei are kept fixed,
we evaluated Eq.~(\ref{eq:ampl_three}), and simulated
the time-dependent build up of the
Fano resonance for the sRICD process
for parameters corresponding to those of the neon dimer at the equilibrium distance of
\unit[3.08]{\AA} \cite{Bondi64} after an excitation to the
Ne$2s^{-1}2p^65p_z$ state. The parameters are given in Table \ref{tab:param}.
All simulations were performed with ELDEST
\cite{ELDEST_v2,SciPy,Numpy1,Numpy2,rapid_input}, where only the outer integral
over $t'$ in Eq.~(\ref{eq:three_compact}) is solved numerically.
We use the Fano parameter $q=10$.
The exact $q$ value is unknown to us, but we expect it to be larger than unity
due to the low probability for direct single photon ionization and excitation.
Since the exact $q$-values are unknown, we assume
equal values for the two continua.
We have performed calculations for a 
range of $q$ parameters, of which a cut is shown in Fig. \ref{fig:nx50_q}.
The time of $t=\unit[93]{fs}$ after the initiating excitation was chosen to
illustrate a typical behaviour.
It becomes evident that the basic features of the spectra are identical, even though
the width and the amplitude
of the signal differ. Therefore, the conclusions of this paper are independent of
the choice of the $q$ parameter.

The exciting pulse for the main results in Figs.~1 and 5
has $A_{0X}=\unit[5\times 10^8]{W/cm^2}$, $\Omega= \unit[47.6930]{eV}$,
and a $\text{FWHM}=\unit[6.1]{fs}$ corresponding to 50 cycles. This duration
ensures that the bandwidth is so small that only the Ne$2s^{-1}2p^65p_z$ state
in the neon dimer is resonantly excited.

We assumed a complete population transfer from the resonant state
of the sRICD process to the resonant state of the ICD process during the second ionizing pulse
over a time of \unit[15]{fs} as realized experimentally\cite{Kaldun16}.

\section{Results and Discussion}

In the investigation of decay processes with short laser pulses, the shortness comes with 
the cost of an energy broadening.
As a consequence, the kinetic energy spectrum of an sRICD electron is given by
a Fano profile centered around the kinetic energy, that corresponds to the energy of the
resonant state, convoluted with a Gaussian centered around the kinetic energy, that would
correspond to a direct excitation and ionization into the final state.
Here, we show the Voigt profile part inherent to most terms
of the absolute square of Eq. (\ref{eq:three_compact}).
\begin{equation}\label{eq:gauss_lorentz}
 P(E_{\text{kin}},\infty)
  \propto \frac{\exp [  -\sigma_t^2 (E_{\textrm{kin}} + E_{\textrm{fin}} - \Omega)^2 ] }
               {(E_{\textrm{kin}} + E_{\textrm{fin}} - E_R)^2 +
                 \frac{\Gamma_{R}^2}{4} }
\end{equation}

\begin{figure}
 \centering
 \includegraphics[width=\columnwidth]{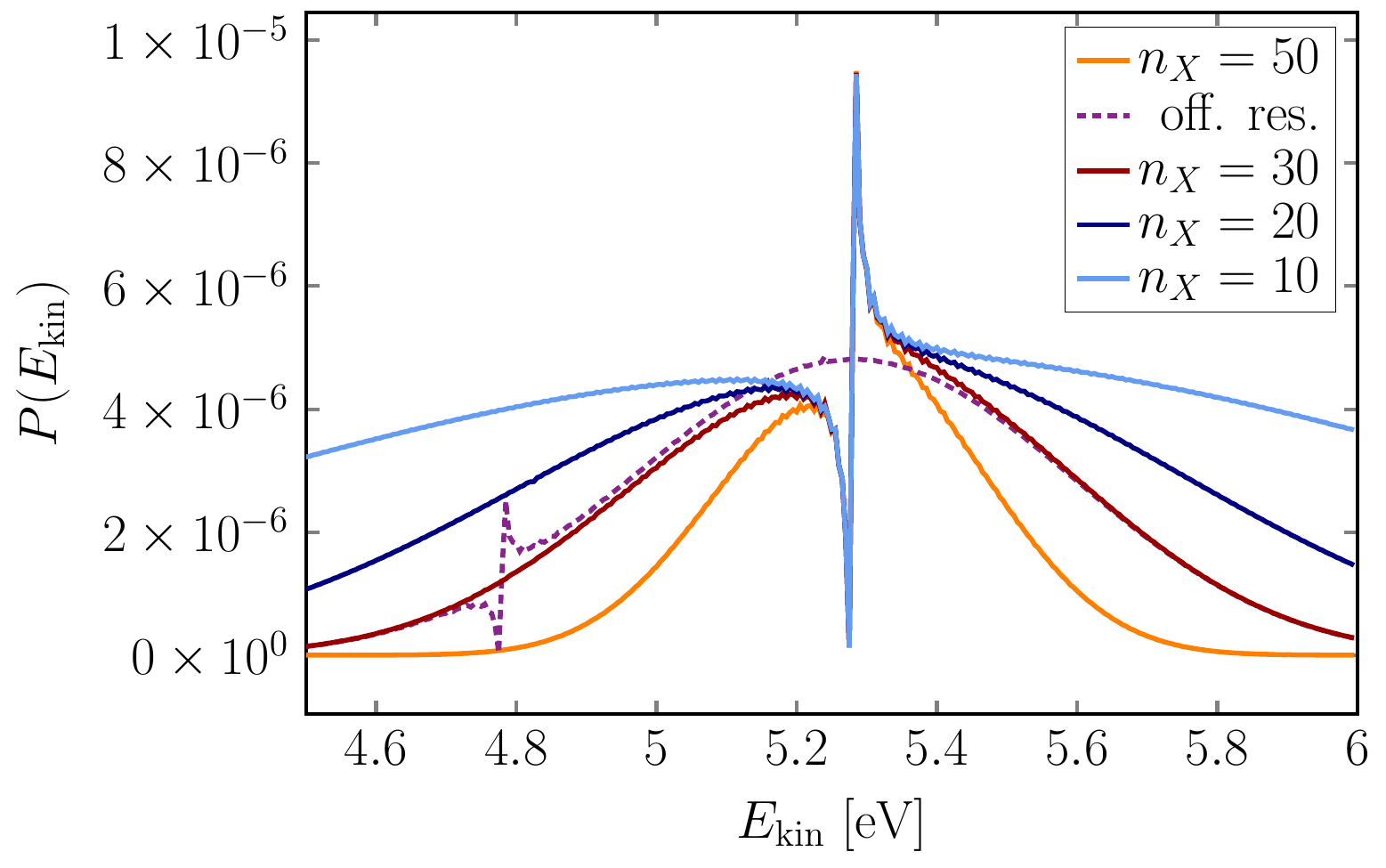}
 \caption{Peak shape of the sRICD electron spectrum for late time
          $t \gg \frac{1}{\Gamma_R}$
          after excitation of our system with electronic parameters corresponding
          to those of the $5p$ 
          state in the neon dimer depending on the number of cycles
          of the exciting pulse $n_X$. The shape can be explained by
          folding the Fourier transform of the exciting pulse in time with a
          Fano profile. In this example, the Fano parameter is  $q=10$.
          Around the resonance energy, the peaks are identical. The width of the
          peak is decreased for an increased number of cycles $n_X$ and
          therefore determined by the duration of the exciting pulse.
          The dashed line illustrates an excitation \unit[0.5]{eV} off resonance for
          $n_X=30$.}
 \label{fig:nx_var}
\end{figure}
The shapes of the spectra are illustrated for different
numbers of cycles $n_X$ for the exciting pulse in Fig.~\ref{fig:nx_var} for a late time
$t \gg \frac{1}{\Gamma_R}$.
There, the mean pulse energy was chosen to be on resonance and therefore, the Fano profile
is centered on the maximum of the Gaussian. 
It is clearly seen, how the peak is narrowed with an increased number of laser
cycles in the exciting pulse.
In order to only excite into a single state
we choose $n_X = 50$, such that other excited states are 
outside $3\sigma_E$ of the energy distribution of the exciting pulse. In future investigations
of the interaction between several decaying resonant states, a smaller number of cycles
will be appropriate.

For the case of an off-resonant excitation, the Fano profile and the Gaussian
will not be centered around the same kinetic energy of the sRICD electron
and the sRICD electron peak will be damped accordingly as
illustrated by the dashed line of Fig. \ref{fig:nx_var} for $n_X=30$.

\begin{figure}
 \centering
 \includegraphics[width=\columnwidth]{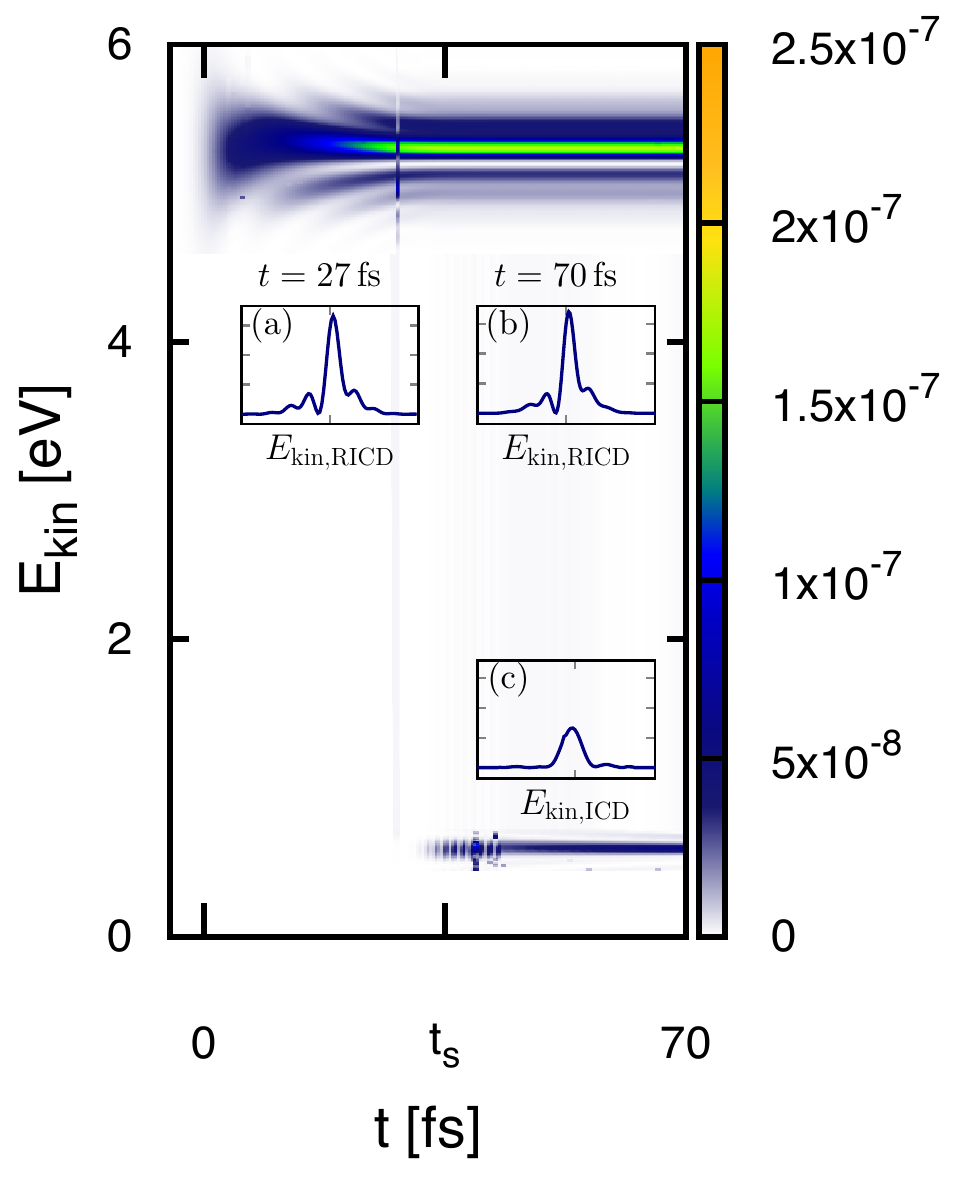}
 \caption{Time-resolved sRICD and ICD spectrum after initial excitation at $t=0$
         and quenching the sRICD at $t=\unit[35]{fs}$.
         The sRICD signal centered around \unit[5.3]{eV} before the second
         pulse shows the buildup of a Fano profile, with
         an additional oscillation in energy
         $\cos\left[ (E_{\mathrm{kin}} + E_{\mathrm{fin}} - E_r) t \right]$
         [see (a) and text]. After
         depopulation of the sRICD resonant state around $t_s$,
         the signal shape remains constant [see (b)].
         For the ICD signal centered around \unit[0.6]{eV} shown in (c),
         only the resonant
         term of Eq.~(\ref{eq:ampl_three}) contributes, which has the shape of a Voigt
         profile. A pump-probe spectrum can be obtained by varying $t_s$
         }
 \label{fig:two_fin_spec}
\end{figure}

By using the parameters for the neon dimer (see Table \ref{tab:param}),
we simulated the time-evolution of
an intial excitation at $t=0$ and quenching of the decay processes by ionization
with a second laser pulse centered at the time $t= \unit[35]{fs}$.
The results are presented in
Fig. \ref{fig:two_fin_spec}. We will discuss its different characteristics separately.

The time-resolved spectrum shows two energy-separated signals within the
chosen kinetic energy
range. The signal of the sRICD process is initiated with the first and the signal of the
ICD process by the second laser pulse, which at the same time reduces
those contributions
of the sRICD signal, that involve a population of the resonant state.
Due to the fast depopulation of the resonant state, the signal before and
after the second pulse are effectively the same
[compare Fig. \ref{fig:two_fin_spec}(a) and (b)]. This allows to generate
a full pump-probe spectrum.
The competing autoionization and pRICD process would result in much higher kinetic electron
energies of about \unit[26]{eV} and can therefore unambigously be distinguished from the
sRICD and ICD electrons shown in Fig.~\ref{fig:two_fin_spec}.

While the sRICD signal shows the buildup of a Fano profile
[see Fig.~\ref{fig:two_fin_spec}(a)], and therefore interference
effects, the ICD signal is an interference free Voigt profile
[see Fig.~\ref{fig:two_fin_spec}(c)],
because only the resonant term [see Eq.~(\ref{eq:ampl_ICD})] significantly
contributes to it. This can be explained by the initiating process being ionization rather
than excitation. In case of the ICD process, the final state is characterized by a doubly
ionized system. A single photon double ionization with low laser intensities is very
unlikely compared to a single ionization. Hence, the corresponding Fano parameter
would be very large and interference effects would have low amplitudes.
At the same time, the kinetic energy distribution
of the two emitted electrons can be expected to be wide and not necessarily
to peak at the energy of the ICD electrons, which would further marginalize the contribution
of the direct pathway. As a result, the signal shape is described by the absolute
square of Eq.~(\ref{eq:ampl_ICD}), i.e., the terms of
Eq. (\ref{eq:res_indir_expl}).

\begin{figure}
 \centering
 \includegraphics[width=\columnwidth]{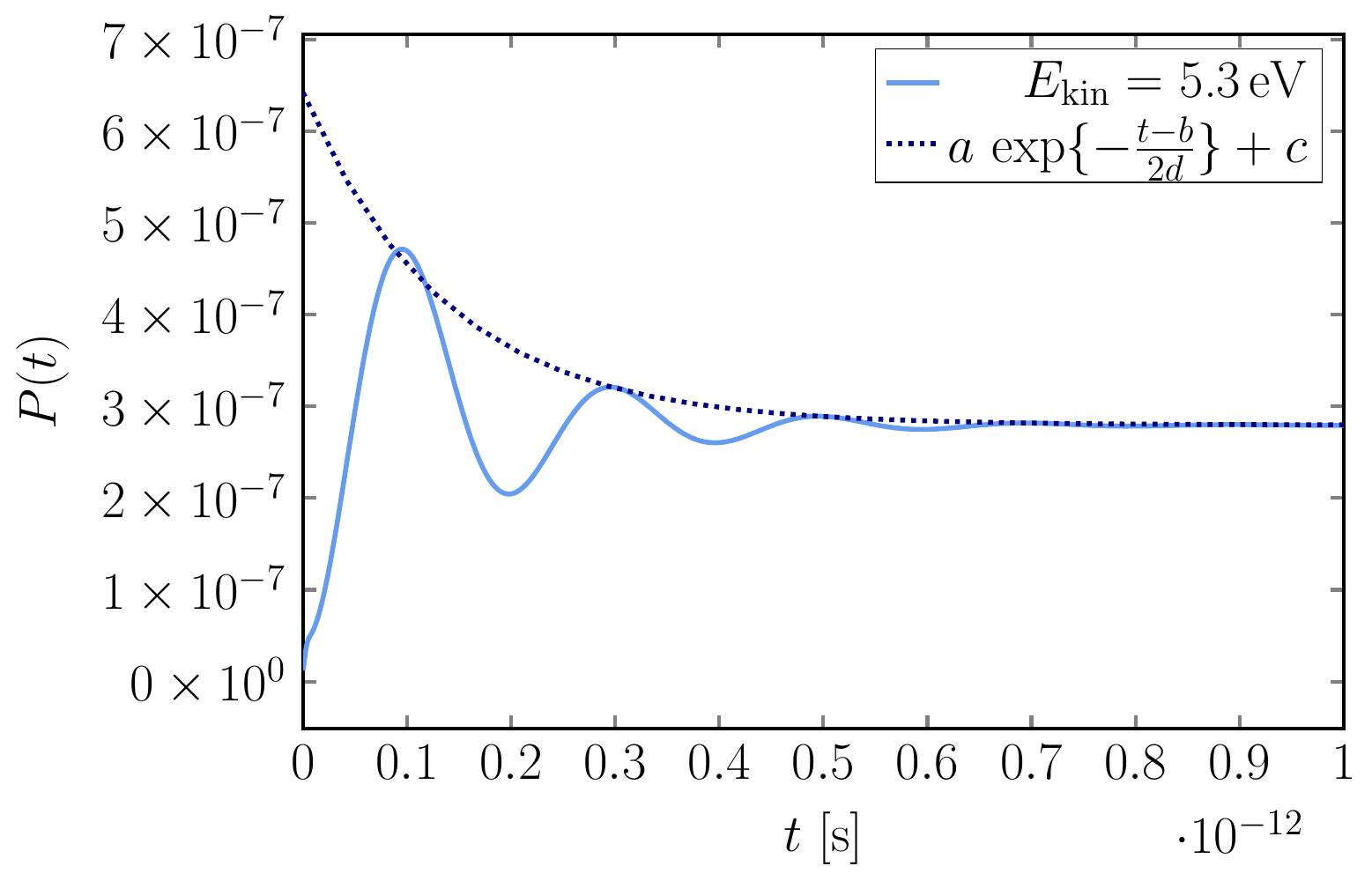}
 \caption{Oscillation in time for observing the sRICD electron with a kinetic energy of
  \unit[5.3]{eV}, for $n_X=50$ and $q=10$. This oscillation
         can be described by
          $\cos\left[ (E_{\mathrm{kin}} + E_{\mathrm{fin}} - E_r) t \right]$.
          Its period is therefore depending on the kinetic energy of the electron.
          It is damped
          in time by the decay of the resonant state $\exp ( - \Gamma_r t /2 )$.
          While the oscillations in energy are known and have been observed in energy,
          the oscillations in time require a process with a long enough lifetime to
          be observed around the resonance energy.
         }
 \label{fig:time_5.3}
\end{figure}

We now turn our attention from the variation of the signals
in energy to their variation in time.
During the build up of the signals over time, they show oscillations in both
energy and time, which are easiest seen for the ICD
in the video supplementing this paper
\cite{video}.
For the sRICD process, the variation in time can be obtained by
solving the dominant
part of the absolute square of Eq.~(\ref{eq:three_compact}) analytically.
For the ICD process, the
dominant part of the absolute square of Eq.~(\ref{eq:ampl_ICD}) needs to be solved.In both cases,
it can be shown that the signal is proportional to
\begin{equation} \label{eq:oscillation}
 \cos [ (E_{r} - E_{\textrm{kin}} - E_{\textrm{fin}}) t ] \,
 \exp ( -\frac{\Gamma_{r}}{2} t ).
\end{equation}
This equation is not specific to the decay process and therefore valid for both
the sRICD and the ICD signal ($r=R,I$).
The variation in energy, illustrated in Fig.~\ref{fig:two_fin_spec}(a), is
in agreement with Ref.~\cite{Wickenhauser_thesis}.
As can be seen from Eq.~(\ref{eq:oscillation}), the oscillation is also visible in time
as shown in Fig.~\ref{fig:time_5.3} for a fixed kinetic energy of the sRICD electron
of $E_{\text{kin}} = \unit[5.3]{eV}$.

In order to be observed, such an oscillation requires long enough lifetimes
of the resonant state.
The comparably slow sRICD process offers the opportunity to observe and utilize it.
An oscillation in time has recently been observed experimentally
for an AI process \cite{Busto18}.
In both the latter and in our case, a discrepancy between the actual and the fitted lifetime
is observed.
Given the analytical expressions of this work, we are able to understand this
discrepancy. 
The kinetic energy in Fig.~\ref{fig:time_5.3} is slightly higher
than the kinetic energy
corresponding to the resonant state. The further away from the resonance this energy is
chosen, the shorter the period of this oscillation.
The oscillation is exponentially damped by the lifetime of the decaying state.
However, the dynamics of the system are more complex than shown
in Eq.~(\ref{eq:oscillation}), because some terms of the resonant and indirect
contributions are damped by $\exp(-\Gamma_r t)$
while others are damped by $\exp(-\Gamma_r t/2)$, which can be seen by evaluating the
time-dependence of the absolute square of Eq. (\ref{eq:three_compact}).
This behaviour is effectively only
observed in the beginning of the decay, because its contribution becomes small within
a short time.
Therefore, a fit of the maxima
at longer times, where the parts damped by $\exp(-\Gamma_r t)$ can be neglected,
to the function $a \exp[-(t-b)/2d] + c$ gives a good estimate of the lifetime
$d\approx \tau_r$. The fit in Fig. \ref{fig:time_5.3}, where the first maximum was
omitted, yields a lifetime of \unit[68]{fs}.
The effective overall lifetime of the excited resonant state used as input parameter
for the simulations is given by
$\frac{1}{\tau_\text{eff}}
 = \frac{1}{\tau_\text{sRICD}} + \frac{1}{\tau_o}$ and $ \tau_\text{eff}= \unit[70]{fs}$
using the parameters given in Table \ref{tab:param}.
The value determined from the fit is slightly lower than this effective lifetime but
still matches it very well.
This kind of fit can therefore be used to determine the lifetime of the electronic
decay process.

\section{Summary}
We have presented a time-dependent theoretical description of the sRICD and ICD processes
investigated with short laser pulses for a model system with frozen nuclei.
We suggest sRICD as the primary process.
We then imagine that it is possible to quench the sRICD process by a strong laser
pulse and thereby initiate the ICD process interference free.
We have shown an oscillation of the ionization probability
in time, which is general for eletronic decay processes. For processes with comparably long
lifetimes, such as the ICD processes, this oscillation allows for a new way of measuring the
lifetime of the underlying processes.
Moreover, the basic formulation opens the door for future investigations in
controlling the decay processes and to guide time-resolved experimental studies.


\section{Acknowledgements}
The research of E.F. was supported by the Villum Foundation.
The research of L.B.M. was supported by the Villum Kann Rasmussen Centre of Excellence
QUSCOPE - Quantum Scale Optical Processes.

\appendix
\section{Fano matrix elements}
\label{sec:app_fano}
\subsection{One continuum final state}
The time-independent wavefunction $\ket{\Psi_{\underline{E}}}$
is a solution to the same Hamiltonian as considered in
Ref. \cite{Fano61}. We therefore have the same solution
\begin{equation}
 \ket{\Psi_{\underline{E}}} = a(\underline{E}) \ket{r}
                              + \int \mathrm{d}E' b_{E'}(\underline{E}) \ket{E'}
\end{equation}
with coefficients
\begin{align}
 a(\underline{E})      =& - \frac{V_r}{\sqrt{(\underline{E}-E_r- F(\underline{E}))^2 + \pi^2 V_r^4} } \\
 b_{E'}(\underline{E}) =& \frac{V_r a(\underline{E})}{\underline{E} - E'} \nonumber \\
                        & - \frac{\underline{E}-E_r - F(\underline{E})}
                             {\sqrt{(\underline{E}-E_r- F(\underline{E}))^2 + \pi^2 V_r^4} }
                             \delta(\underline{E} - E') .
\end{align}
Here, $F(\underline{E})$ is a small shift in the energy position of the
resonant state $\ket{r}$,
which is of second order in $V$ and, which we neglect in the following.

The Fano matrix elements of Eq.~(\ref{eq:ampl_three}) involving the Fano time-evolution
operator can now be solved by contour integration in the negative
complex half-plane, where $a_{-1}$ is the first order residue and we use
that $\braket{r|r}=1$, $\braket{E|E'}=\delta(E-E')$ and $\braket{E|r}=0$:

\begin{widetext}
\begin{align}
 \braket{r| U_F(t'',t')| r}
 =& \int\mathrm{d}\underline{E} \,\,  \exp[-i\underline{E} (t''-t') ] \, |a(\underline{E})|^2 \\
 =& V_r^2 \int\mathrm{d}\underline{E} \,\, \frac{ \exp[-i\underline{E} (t''-t')] }
                                             {(\underline{E} - E_r)^2 + \pi^2 V_r^4} \\
 =& V_r^2 \int\mathrm{d}\underline{E} \,\, \frac{ \exp[-i\underline{E} (t''-t')] }
           {(\underline{E} - E_r + i\pi V_r^2) \, (\underline{E} - E_r - i\pi V_r^2)} \\
 =& V_r^2 \, \exp[ -iE_r (t''-t') ]
    \int\mathrm{d}\tilde{E} \,\, \frac{ \exp[-i\tilde{E} (t''-t')] }
           {(\tilde{E} + i\pi V_r^2) \, (\tilde{E} - i\pi V_r^2)} \\
 =& 2\pi i \, V_r^2 \, \exp[ -iE_r (t''-t') ] \, a_{-1} \\
 =& \exp[ -i (E_r -i \pi V_r^2) (t'' - t')]
\end{align}

\begin{align}
 \int \mathrm{d}E' \braket{r| U_F(t'',t')| E'}
 =& \int\mathrm{d}\underline{E} \int\mathrm{d}E' \,\,  \exp[-i\underline{E} (t''-t')] \,
                a(\underline{E}) b_{E'}^*(\underline{E}) \\
 =& V_r^2 \int\mathrm{d}\underline{E} \,\, \frac{ (\underline{E} - E_r)
                                                 \exp[-i\underline{E} (t''-t')] }
                                             {(\underline{E} - E_r)^2 + \pi^2 V_r^4} \\
 =& V_r^2 \int\mathrm{d}\underline{E} \,\, \frac{ (\underline{E} - E_r)
                                                  \exp[-i\underline{E} (t''-t')] }
           {(\underline{E} - E_r + i\pi V_r^2) \, (\underline{E} - E_r - i\pi V_r^2)} \\
 =& V_r^2 \, \exp[ -iE_r (t''-t') ] 
           \int\mathrm{d}\tilde{E} \,\, \frac{ \tilde{E} \exp[-i\tilde{E} (t''-t')] }
           {(\tilde{E} + i\pi V_r^2) \, (\tilde{E} - i\pi V_r^2)} \\
 =& 2\pi i \, V_r^2 \, \exp[ -iE_r (t''-t') ] \, a_{-1} \\
 =& -i \pi V_r \exp[ -i (E_r -i \pi V_r^2) (t'' - t')]
\end{align}
\end{widetext}

\subsection{Two continuum final states}
The time-independent wavefunction $\ket{\Psi_{\underline{E}}}$
is a solution to the same Hamiltonian as considered in
Ref. \cite{Fano61}. After neglecting $F(\underline{E})$ as before we therefore
have the same solutions
\begin{equation}
 \ket{\Psi_{\underline{E}}} = a(\underline{E}) \ket{r}
                              + \int \mathrm{d}E_1' b_{E_1'}(\underline{E}) \ket{E_1'}
                              + \int \mathrm{d}E_2' b_{E_2'}(\underline{E}) \ket{E_2'}
\end{equation}
with coefficients
\begin{widetext}
\begin{align}
 a(\underline{E})      =& - \frac{\sqrt{V_r^2 + W_r^2}}
                            {\sqrt{(\underline{E}-E_r)^2 + \pi^2 (V_r^2 + W_r^2)^2} } \\
 b_{E_1'}(\underline{E}) =& \frac{V_r a(\underline{E})}{\sqrt{V_r^2 + W_r^2} (\underline{E} - E_1')}
                          - \frac{\underline{E}-E_r}
                             {\sqrt{(\underline{E}-E_r)^2 + \pi^2 (V_r^2 + W_r^2)^2} }
                             \delta(\underline{E} - E_1') \\
 c_{E_2'}(\underline{E}) =& \frac{W_r a(\underline{E})}{\sqrt{V_r^2 + W_r^2} (\underline{E} - E_2')}
                          - \frac{\underline{E}-E_r}
                             {\sqrt{(\underline{E}-E_r)^2 + \pi^2 (V_r^2 + W_r^2)^2} }
                             \delta(\underline{E} - E_2') .
\end{align}
\end{widetext}

The Fano integrals can now be solved by contour integration in the negative
complex half-plane, where $a_{-1}$ is the first order residue and we use
that $\braket{r|r}=1$, $\braket{E|E'}=\delta(E-E')$ and $\braket{E|r}=0$:
\begin{widetext}
\begin{align}
 \braket{r| U_F(t'',t')| r}
 =& (V_r^2 + W_r^2) \int\mathrm{d}\underline{E} \,\,  \exp[-i\underline{E} (t''-t') ] \, |a(\underline{E})|^2 \\
 =& (V_r^2 + W_r^2) \int\mathrm{d}\underline{E} \,\, \frac{ \exp[-i\underline{E} (t''-t')] }
                                             {(\underline{E} - E_r)^2 + \pi^2 (V_r^2 + W_r^2)^2} \\
 =& (V_r^2 + W_r^2) \int\mathrm{d}\underline{E} \,\, \frac{ \exp[-i\underline{E} (t''-t')] }
           {(\underline{E} - E_r + i\pi (V_r^2 + W_r^2)) \, (\underline{E} - E_r - i\pi (V_r^2 + W_r^2))} \\
 =& (V_r^2 + W_r^2) \, \exp[ -iE_r (t''-t') ]
    \int\mathrm{d}\tilde{E} \,\, \frac{ \exp[-i\tilde{E} (t''-t')] }
           {(\tilde{E} + i\pi (V_r^2 + W_r^2)) \, (\tilde{E} - i\pi (V_r^2 + W_r^2))} \\
 =& 2\pi i \, (V_r^2 + W_r^2) \, \exp[ -iE_r (t''-t') ] \, a_{-1} \\
 =& \exp[ -i (E_r -i \pi (V_r^2 + W_r^2)) (t'' - t')]
\end{align}

\begin{align}
 \int \mathrm{d}E' \braket{r| U_F(t'',t')| E'}
 =& \int\mathrm{d}\underline{E} \int\mathrm{d}E' \,\,  \exp[-i\underline{E} (t''-t')] \,
                a(\underline{E}) b_{E'}^*(\underline{E}) \\
 =& V_r \int\mathrm{d}\underline{E} \,\, \frac{ (\underline{E} - E_r)
                                                 \exp[-i\underline{E} (t''-t')] }
                                             {(\underline{E} - E_r)^2 + \pi^2 (V_r^2 + W_r^2)^2} \\
 =& V_r \int\mathrm{d}\underline{E} \,\, \frac{ (\underline{E} - E_r)
                                                  \exp[-i\underline{E} (t''-t')] }
           {(\underline{E} - E_r + i\pi (V_r^2 + W_r^2)) \, (\underline{E} - E_r - i\pi (V_r^2 + W_r^2))} \\
 =& V_r \, \exp[ -iE_r (t''-t') ] 
           \int\mathrm{d}\tilde{E} \,\, \frac{ \tilde{E} \exp[-i\tilde{E} (t''-t')] }
           {(\tilde{E} + i\pi (V_r^2 + W_r^2)) \, (\tilde{E} - i\pi (V_r^2 + W_r^2))} \\
 =& 2\pi i \, V_r \, \exp[ -iE_r (t''-t') ] \, a_{-1} \\
 =& -i \pi V_r \exp[ -i (E_r -i \pi (V_r^2 + W_r^2)) (t'' - t')] .
\end{align}
\end{widetext}

\bibliography{theolit}

\begin{thebibliography}{55}%
\makeatletter
\providecommand \@ifxundefined [1]{%
 \@ifx{#1\undefined}
}%
\providecommand \@ifnum [1]{%
 \ifnum #1\expandafter \@firstoftwo
 \else \expandafter \@secondoftwo
 \fi
}%
\providecommand \@ifx [1]{%
 \ifx #1\expandafter \@firstoftwo
 \else \expandafter \@secondoftwo
 \fi
}%
\providecommand \natexlab [1]{#1}%
\providecommand \enquote  [1]{``#1''}%
\providecommand \bibnamefont  [1]{#1}%
\providecommand \bibfnamefont [1]{#1}%
\providecommand \citenamefont [1]{#1}%
\providecommand \href@noop [0]{\@secondoftwo}%
\providecommand \href [0]{\begingroup \@sanitize@url \@href}%
\providecommand \@href[1]{\@@startlink{#1}\@@href}%
\providecommand \@@href[1]{\endgroup#1\@@endlink}%
\providecommand \@sanitize@url [0]{\catcode `\\12\catcode `\$12\catcode
  `\&12\catcode `\#12\catcode `\^12\catcode `\_12\catcode `\%12\relax}%
\providecommand \@@startlink[1]{}%
\providecommand \@@endlink[0]{}%
\providecommand \url  [0]{\begingroup\@sanitize@url \@url }%
\providecommand \@url [1]{\endgroup\@href {#1}{\urlprefix }}%
\providecommand \urlprefix  [0]{URL }%
\providecommand \Eprint [0]{\href }%
\providecommand \doibase [0]{http://dx.doi.org/}%
\providecommand \selectlanguage [0]{\@gobble}%
\providecommand \bibinfo  [0]{\@secondoftwo}%
\providecommand \bibfield  [0]{\@secondoftwo}%
\providecommand \translation [1]{[#1]}%
\providecommand \BibitemOpen [0]{}%
\providecommand \bibitemStop [0]{}%
\providecommand \bibitemNoStop [0]{.\EOS\space}%
\providecommand \EOS [0]{\spacefactor3000\relax}%
\providecommand \BibitemShut  [1]{\csname bibitem#1\endcsname}%
\let\auto@bib@innerbib\@empty
\bibitem [{\citenamefont {Cederbaum}\ \emph {et~al.}(1997)\citenamefont
  {Cederbaum}, \citenamefont {Zobeley},\ and\ \citenamefont
  {Tarantelli}}]{Cederbaum97}%
  \BibitemOpen
  \bibfield  {author} {\bibinfo {author} {\bibfnamefont {L.~S.}\ \bibnamefont
  {Cederbaum}}, \bibinfo {author} {\bibfnamefont {J.}~\bibnamefont {Zobeley}},
  \ and\ \bibinfo {author} {\bibfnamefont {F.}~\bibnamefont {Tarantelli}},\
  }\bibfield  {title} {\enquote {\bibinfo {title} {{Giant Intermolecular Decay
  and Fragmentation of Clusters}},}\ }\href@noop {} {\bibfield  {journal}
  {\bibinfo  {journal} {Phys. Rev. Lett.}\ }\textbf {\bibinfo {volume} {79}},\
  \bibinfo {pages} {4778} (\bibinfo {year} {1997})}\BibitemShut {NoStop}%
\bibitem [{\citenamefont {Marburger}\ \emph {et~al.}(2003)\citenamefont
  {Marburger}, \citenamefont {Kugeler}, \citenamefont {Hergenhahn},\ and\
  \citenamefont {M{\"o}ller}}]{Marburger03}%
  \BibitemOpen
  \bibfield  {author} {\bibinfo {author} {\bibfnamefont {S.}~\bibnamefont
  {Marburger}}, \bibinfo {author} {\bibfnamefont {O.}~\bibnamefont {Kugeler}},
  \bibinfo {author} {\bibfnamefont {U.}~\bibnamefont {Hergenhahn}}, \ and\
  \bibinfo {author} {\bibfnamefont {T.}~\bibnamefont {M{\"o}ller}},\ }\bibfield
   {title} {\enquote {\bibinfo {title} {Experimental evidence for interatomic
  coulombic decay in ne clusters},}\ }\href@noop {} {\bibfield  {journal}
  {\bibinfo  {journal} {Phys. Rev. Lett.}\ }\textbf {\bibinfo {volume} {90}},\
  \bibinfo {pages} {203401} (\bibinfo {year} {2003})}\BibitemShut {NoStop}%
\bibitem [{\citenamefont {Santra}\ \emph {et~al.}(2000)\citenamefont {Santra},
  \citenamefont {Zobeley}, \citenamefont {Cederbaum},\ and\ \citenamefont
  {Moiseyev}}]{Santra00_1}%
  \BibitemOpen
  \bibfield  {author} {\bibinfo {author} {\bibfnamefont {R.}~\bibnamefont
  {Santra}}, \bibinfo {author} {\bibfnamefont {J.}~\bibnamefont {Zobeley}},
  \bibinfo {author} {\bibfnamefont {L.~S.}\ \bibnamefont {Cederbaum}}, \ and\
  \bibinfo {author} {\bibfnamefont {N.}~\bibnamefont {Moiseyev}},\ }\bibfield
  {title} {\enquote {\bibinfo {title} {{Interatomic Coulombic Decay in van der
  Waals Clusters and Impact of Nuclear Motion}},}\ }\href@noop {} {\bibfield
  {journal} {\bibinfo  {journal} {Phys. Rev. Lett.}\ }\textbf {\bibinfo
  {volume} {85}},\ \bibinfo {pages} {4490} (\bibinfo {year}
  {2000})}\BibitemShut {NoStop}%
\bibitem [{\citenamefont {Hergenhahn}(2011)}]{Hergenhahn11}%
  \BibitemOpen
  \bibfield  {author} {\bibinfo {author} {\bibfnamefont {U.}~\bibnamefont
  {Hergenhahn}},\ }\bibfield  {title} {\enquote {\bibinfo {title} {{Interatomic
  and Intermolecular Coulombic Decay: The Early Years}},}\ }\href@noop {}
  {\bibfield  {journal} {\bibinfo  {journal} {J. Electron Spectrosc. Relat.
  Phenom.}\ }\textbf {\bibinfo {volume} {184}},\ \bibinfo {pages} {78}
  (\bibinfo {year} {2011})}\BibitemShut {NoStop}%
\bibitem [{\citenamefont {Jahnke}(2015)}]{Jahnke15}%
  \BibitemOpen
  \bibfield  {author} {\bibinfo {author} {\bibfnamefont {T.}~\bibnamefont
  {Jahnke}},\ }\bibfield  {title} {\enquote {\bibinfo {title} {{Interatomic and
  intermolecular Coulombic decay: the coming of age story}},}\ }\href@noop {}
  {\bibfield  {journal} {\bibinfo  {journal} {J. Phys. B: Atomic, Molecular and
  Optical Physics}\ }\textbf {\bibinfo {volume} {48}},\ \bibinfo {pages}
  {082001} (\bibinfo {year} {2015})}\BibitemShut {NoStop}%
\bibitem [{\citenamefont {Fasshauer}(2016)}]{Fasshauer16}%
  \BibitemOpen
  \bibfield  {author} {\bibinfo {author} {\bibfnamefont {E.}~\bibnamefont
  {Fasshauer}},\ }\bibfield  {title} {\enquote {\bibinfo {title} {{Non-nearest
  neighbour ICD in clusters}},}\ }\href@noop {} {\bibfield  {journal} {\bibinfo
   {journal} {New J. Phys.}\ }\textbf {\bibinfo {volume} {18}},\ \bibinfo
  {pages} {043028} (\bibinfo {year} {2016})}\BibitemShut {NoStop}%
\bibitem [{\citenamefont {Fa{\ss}hauer}\ \emph {et~al.}(2010)\citenamefont
  {Fa{\ss}hauer}, \citenamefont {Kryzhevoi},\ and\ \citenamefont
  {Pernpointner}}]{Fasshauer10}%
  \BibitemOpen
  \bibfield  {author} {\bibinfo {author} {\bibfnamefont {E.}~\bibnamefont
  {Fa{\ss}hauer}}, \bibinfo {author} {\bibfnamefont {N.~V.}\ \bibnamefont
  {Kryzhevoi}}, \ and\ \bibinfo {author} {\bibfnamefont {M.}~\bibnamefont
  {Pernpointner}},\ }\bibfield  {title} {\enquote {\bibinfo {title} {{Possible
  electronic decay channels in the ionization spectra of small clusters
  composed of Ar and Xe: A four-component relativistic treatment}},}\
  }\href@noop {} {\bibfield  {journal} {\bibinfo  {journal} {J. Chem. Phys.}\
  }\textbf {\bibinfo {volume} {133}},\ \bibinfo {pages} {014303} (\bibinfo
  {year} {2010})}\BibitemShut {NoStop}%
\bibitem [{\citenamefont {Fasshauer}\ \emph {et~al.}(2013)\citenamefont
  {Fasshauer}, \citenamefont {Pernpointner},\ and\ \citenamefont
  {Gokhberg}}]{Fasshauer13}%
  \BibitemOpen
  \bibfield  {author} {\bibinfo {author} {\bibfnamefont {E.}~\bibnamefont
  {Fasshauer}}, \bibinfo {author} {\bibfnamefont {M.}~\bibnamefont
  {Pernpointner}}, \ and\ \bibinfo {author} {\bibfnamefont {K.}~\bibnamefont
  {Gokhberg}},\ }\bibfield  {title} {\enquote {\bibinfo {title} {Interatomic
  decay of inner-valence ionized states in arxe clusters: Relativistic
  approach},}\ }\href@noop {} {\bibfield  {journal} {\bibinfo  {journal} {J.
  Chem. Phys.}\ }\textbf {\bibinfo {volume} {138}},\ \bibinfo {pages} {014305}
  (\bibinfo {year} {2013})}\BibitemShut {NoStop}%
\bibitem [{\citenamefont {Fasshauer}\ \emph {et~al.}(2014)\citenamefont
  {Fasshauer}, \citenamefont {F\"orstel}, \citenamefont {Pallmann},
  \citenamefont {Pernpointner},\ and\ \citenamefont
  {Hergenhahn}}]{Fasshauer14_1}%
  \BibitemOpen
  \bibfield  {author} {\bibinfo {author} {\bibfnamefont {E.}~\bibnamefont
  {Fasshauer}}, \bibinfo {author} {\bibfnamefont {M.}~\bibnamefont
  {F\"orstel}}, \bibinfo {author} {\bibfnamefont {S.}~\bibnamefont {Pallmann}},
  \bibinfo {author} {\bibfnamefont {M.}~\bibnamefont {Pernpointner}}, \ and\
  \bibinfo {author} {\bibfnamefont {U.}~\bibnamefont {Hergenhahn}},\ }\bibfield
   {title} {\enquote {\bibinfo {title} {Using icd for structure analysis of
  clusters: a case study on coexpanded ne-ar clusters},}\ }\href@noop {}
  {\bibfield  {journal} {\bibinfo  {journal} {New J. Phys.}\ }\textbf {\bibinfo
  {volume} {16}},\ \bibinfo {pages} {103026} (\bibinfo {year}
  {2014})}\BibitemShut {NoStop}%
\bibitem [{\citenamefont {Förstel}\ \emph {et~al.}(2016)\citenamefont
  {Förstel}, \citenamefont {Mucke}, \citenamefont {Arion}, \citenamefont
  {Lischke}, \citenamefont {Pernpointner}, \citenamefont {Hergenhahn},\ and\
  \citenamefont {Fasshauer}}]{Foerstel16}%
  \BibitemOpen
  \bibfield  {author} {\bibinfo {author} {\bibfnamefont {M.}~\bibnamefont
  {Förstel}}, \bibinfo {author} {\bibfnamefont {M.}~\bibnamefont {Mucke}},
  \bibinfo {author} {\bibfnamefont {T.}~\bibnamefont {Arion}}, \bibinfo
  {author} {\bibfnamefont {T.}~\bibnamefont {Lischke}}, \bibinfo {author}
  {\bibfnamefont {M.}~\bibnamefont {Pernpointner}}, \bibinfo {author}
  {\bibfnamefont {U.}~\bibnamefont {Hergenhahn}}, \ and\ \bibinfo {author}
  {\bibfnamefont {E}~\bibnamefont {Fasshauer}},\ }\bibfield  {title} {\enquote
  {\bibinfo {title} {{Long-Range Interatomic Coulombic Decay in ArXe Clusters:
  Experiment and Theory}},}\ }\href@noop {} {\bibfield  {journal} {\bibinfo
  {journal} {J. Phys. Chem. C}\ }\textbf {\bibinfo {volume} {120}},\ \bibinfo
  {pages} {22957} (\bibinfo {year} {2016})}\BibitemShut {NoStop}%
\bibitem [{\citenamefont {Fasshauer}\ \emph {et~al.}(2017)\citenamefont
  {Fasshauer}, \citenamefont {Förstel}, \citenamefont {Mucke}, \citenamefont
  {Arion},\ and\ \citenamefont {Hergenhahn}}]{Fasshauer17}%
  \BibitemOpen
  \bibfield  {author} {\bibinfo {author} {\bibfnamefont {E.}~\bibnamefont
  {Fasshauer}}, \bibinfo {author} {\bibfnamefont {M.}~\bibnamefont {Förstel}},
  \bibinfo {author} {\bibfnamefont {M.}~\bibnamefont {Mucke}}, \bibinfo
  {author} {\bibfnamefont {T.}~\bibnamefont {Arion}}, \ and\ \bibinfo {author}
  {\bibfnamefont {U.}~\bibnamefont {Hergenhahn}},\ }\bibfield  {title}
  {\enquote {\bibinfo {title} {{Theoretical and experimental investigation of
  Electron Transfer Mediated Decay in ArKr clusters}},}\ }\href@noop {}
  {\bibfield  {journal} {\bibinfo  {journal} {Chem. Phys.}\ }\textbf {\bibinfo
  {volume} {482}},\ \bibinfo {pages} {226} (\bibinfo {year}
  {2017})}\BibitemShut {NoStop}%
\bibitem [{\citenamefont {M{\"u}ller}\ and\ \citenamefont
  {Cederbaum}(2006)}]{Mueller06}%
  \BibitemOpen
  \bibfield  {author} {\bibinfo {author} {\bibfnamefont {I.~B.}\ \bibnamefont
  {M{\"u}ller}}\ and\ \bibinfo {author} {\bibfnamefont {L.~S.}\ \bibnamefont
  {Cederbaum}},\ }\bibfield  {title} {\enquote {\bibinfo {title} {{Ionization
  and double ionization of small water clusters}},}\ }\href@noop {} {\bibfield
  {journal} {\bibinfo  {journal} {J. Chem. Phys.}\ }\textbf {\bibinfo {volume}
  {125}},\ \bibinfo {pages} {204305} (\bibinfo {year} {2006})}\BibitemShut
  {NoStop}%
\bibitem [{\citenamefont {Kryzhevoi}\ and\ \citenamefont
  {Cederbaum}(2011{\natexlab{a}})}]{Kryzhevoi11_1}%
  \BibitemOpen
  \bibfield  {author} {\bibinfo {author} {\bibfnamefont {N.~V.}\ \bibnamefont
  {Kryzhevoi}}\ and\ \bibinfo {author} {\bibfnamefont {L.~S.}\ \bibnamefont
  {Cederbaum}},\ }\bibfield  {title} {\enquote {\bibinfo {title} {{Using pH
  Value To Control Intermolecular Electronic Decay}},}\ }\href@noop {}
  {\bibfield  {journal} {\bibinfo  {journal} {Angew. Chem. Int. Ed.}\ }\textbf
  {\bibinfo {volume} {50}},\ \bibinfo {pages} {1306} (\bibinfo {year}
  {2011}{\natexlab{a}})}\BibitemShut {NoStop}%
\bibitem [{\citenamefont {Stoychev}\ \emph {et~al.}(2011)\citenamefont
  {Stoychev}, \citenamefont {Kuleff},\ and\ \citenamefont
  {Cederbaum}}]{Stoychev11}%
  \BibitemOpen
  \bibfield  {author} {\bibinfo {author} {\bibfnamefont {S.~D.}\ \bibnamefont
  {Stoychev}}, \bibinfo {author} {\bibfnamefont {A.~I.}\ \bibnamefont
  {Kuleff}}, \ and\ \bibinfo {author} {\bibfnamefont {L.~S.}\ \bibnamefont
  {Cederbaum}},\ }\bibfield  {title} {\enquote {\bibinfo {title}
  {{Intermolecular Coulombic Decay in Small Biochemically Relevant
  Hydrogen-Bonded Systems}},}\ }\href@noop {} {\bibfield  {journal} {\bibinfo
  {journal} {J. Am. Chem. Soc.}\ }\textbf {\bibinfo {volume} {133}},\ \bibinfo
  {pages} {6817} (\bibinfo {year} {2011})}\BibitemShut {NoStop}%
\bibitem [{\citenamefont {Kryzhevoi}\ and\ \citenamefont
  {Cederbaum}(2011{\natexlab{b}})}]{Kryzhevoi11_2}%
  \BibitemOpen
  \bibfield  {author} {\bibinfo {author} {\bibfnamefont {N.~V.}\ \bibnamefont
  {Kryzhevoi}}\ and\ \bibinfo {author} {\bibfnamefont {L.~S.}\ \bibnamefont
  {Cederbaum}},\ }\bibfield  {title} {\enquote {\bibinfo {title} {{Nonlocal
  Effects in the Core Ionization and Auger Spectra of Small Ammonia
  Clusters}},}\ }\href {\doibase http://dx.doi.org/10.1021/jp109920p}
  {\bibfield  {journal} {\bibinfo  {journal} {J. Phys. Chem. B}\ }\textbf
  {\bibinfo {volume} {115}},\ \bibinfo {pages} {5441} (\bibinfo {year}
  {2011}{\natexlab{b}})}\BibitemShut {NoStop}%
\bibitem [{\citenamefont {Oostenrijk}\ \emph {et~al.}(2018)\citenamefont
  {Oostenrijk}, \citenamefont {Walsh}, \citenamefont {Laksman}, \citenamefont
  {Mansson}, \citenamefont {Grunewald}, \citenamefont {Sorensen},\ and\
  \citenamefont {Gisselbrecht}}]{Oostenrijk18}%
  \BibitemOpen
  \bibfield  {author} {\bibinfo {author} {\bibfnamefont {B.}~\bibnamefont
  {Oostenrijk}}, \bibinfo {author} {\bibfnamefont {N.}~\bibnamefont {Walsh}},
  \bibinfo {author} {\bibfnamefont {J.}~\bibnamefont {Laksman}}, \bibinfo
  {author} {\bibfnamefont {E.~P.}\ \bibnamefont {Mansson}}, \bibinfo {author}
  {\bibfnamefont {Ch.}\ \bibnamefont {Grunewald}}, \bibinfo {author}
  {\bibfnamefont {S.~L.}\ \bibnamefont {Sorensen}}, \ and\ \bibinfo {author}
  {\bibfnamefont {M.}~\bibnamefont {Gisselbrecht}},\ }\bibfield  {title}
  {\enquote {\bibinfo {title} {{The role of charge and proton transfer in
  fragmentation of hydrogen-bonded nanosystems: the breakup of ammonia clusters
  upon single photon multi-ionization}},}\ }\href@noop {} {\bibfield  {journal}
  {\bibinfo  {journal} {Phys. Chem. Chem. Phys.}\ }\textbf {\bibinfo {volume}
  {20}},\ \bibinfo {pages} {932} (\bibinfo {year} {2018})}\BibitemShut
  {NoStop}%
\bibitem [{\citenamefont {Harbach}\ \emph {et~al.}(2013)\citenamefont
  {Harbach}, \citenamefont {Schneider}, \citenamefont {Faraji},\ and\
  \citenamefont {Dreuw}}]{Harbach13}%
  \BibitemOpen
  \bibfield  {author} {\bibinfo {author} {\bibfnamefont {P.~H.~P.}\
  \bibnamefont {Harbach}}, \bibinfo {author} {\bibfnamefont {M.}~\bibnamefont
  {Schneider}}, \bibinfo {author} {\bibfnamefont {S.}~\bibnamefont {Faraji}}, \
  and\ \bibinfo {author} {\bibfnamefont {A.}~\bibnamefont {Dreuw}},\ }\bibfield
   {title} {\enquote {\bibinfo {title} {Intermolecular coulombic decay in
  biology: the initial electron detachment from fadh- in dna photolyases},}\
  }\href@noop {} {\bibfield  {journal} {\bibinfo  {journal} {J. Phys. Chem.
  Lett.}\ }\textbf {\bibinfo {volume} {4}},\ \bibinfo {pages} {943} (\bibinfo
  {year} {2013})}\BibitemShut {NoStop}%
\bibitem [{\citenamefont {Stumpf}\ \emph {et~al.}(2016)\citenamefont {Stumpf},
  \citenamefont {Gokhberg},\ and\ \citenamefont {Cederbaum}}]{Stumpf16a}%
  \BibitemOpen
  \bibfield  {author} {\bibinfo {author} {\bibfnamefont {V.}~\bibnamefont
  {Stumpf}}, \bibinfo {author} {\bibfnamefont {K.}~\bibnamefont {Gokhberg}}, \
  and\ \bibinfo {author} {\bibfnamefont {L.~S.}\ \bibnamefont {Cederbaum}},\
  }\bibfield  {title} {\enquote {\bibinfo {title} {{The role of metal ions in
  X-ray-induced photochemistry}},}\ }\href@noop {} {\bibfield  {journal}
  {\bibinfo  {journal} {Nature Chemistry}\ }\textbf {\bibinfo {volume} {8}},\
  \bibinfo {pages} {237} (\bibinfo {year} {2016})}\BibitemShut {NoStop}%
\bibitem [{\citenamefont {Bouda\"iffa}\ \emph {et~al.}(2000)\citenamefont
  {Bouda\"iffa}, \citenamefont {Cloutier}, \citenamefont {Hunting},
  \citenamefont {Huels},\ and\ \citenamefont {Sanche}}]{Boudaiffa00}%
  \BibitemOpen
  \bibfield  {author} {\bibinfo {author} {\bibfnamefont {Badia}\ \bibnamefont
  {Bouda\"iffa}}, \bibinfo {author} {\bibfnamefont {Pierre}\ \bibnamefont
  {Cloutier}}, \bibinfo {author} {\bibfnamefont {Darel}\ \bibnamefont
  {Hunting}}, \bibinfo {author} {\bibfnamefont {Michael~A.}\ \bibnamefont
  {Huels}}, \ and\ \bibinfo {author} {\bibfnamefont {L\'eon}\ \bibnamefont
  {Sanche}},\ }\bibfield  {title} {\enquote {\bibinfo {title} {Resonant
  formation of dna strand breaks by low-energy (3 to 20 ev) electrons},}\
  }\href@noop {} {\bibfield  {journal} {\bibinfo  {journal} {Science}\ }\textbf
  {\bibinfo {volume} {287}},\ \bibinfo {pages} {1658--1660} (\bibinfo {year}
  {2000})}\BibitemShut {NoStop}%
\bibitem [{\citenamefont {Brun}\ \emph {et~al.}(2009)\citenamefont {Brun},
  \citenamefont {Cloutier}, \citenamefont {Sicard-Roselli}, \citenamefont
  {Fromm},\ and\ \citenamefont {Sanche}}]{Brun09}%
  \BibitemOpen
  \bibfield  {author} {\bibinfo {author} {\bibfnamefont {E.}~\bibnamefont
  {Brun}}, \bibinfo {author} {\bibfnamefont {P.}~\bibnamefont {Cloutier}},
  \bibinfo {author} {\bibfnamefont {C.}~\bibnamefont {Sicard-Roselli}},
  \bibinfo {author} {\bibfnamefont {M.}~\bibnamefont {Fromm}}, \ and\ \bibinfo
  {author} {\bibfnamefont {L.}~\bibnamefont {Sanche}},\ }\bibfield  {title}
  {\enquote {\bibinfo {title} {Damage induced to dna by low-energy (0--30 ev)
  electrons under vacuum and atmospheric conditions},}\ }\href@noop {}
  {\bibfield  {journal} {\bibinfo  {journal} {J. Phys. Chem. B}\ }\textbf
  {\bibinfo {volume} {113}},\ \bibinfo {pages} {10008} (\bibinfo {year}
  {2009})}\BibitemShut {NoStop}%
\bibitem [{\citenamefont {Pan}\ \emph {et~al.}(2003)\citenamefont {Pan},
  \citenamefont {Cloutier}, \citenamefont {Hunting},\ and\ \citenamefont
  {Sanche}}]{Pan03}%
  \BibitemOpen
  \bibfield  {author} {\bibinfo {author} {\bibfnamefont {X.}~\bibnamefont
  {Pan}}, \bibinfo {author} {\bibfnamefont {P.}~\bibnamefont {Cloutier}},
  \bibinfo {author} {\bibfnamefont {D.}~\bibnamefont {Hunting}}, \ and\
  \bibinfo {author} {\bibfnamefont {L.}~\bibnamefont {Sanche}},\ }\bibfield
  {title} {\enquote {\bibinfo {title} {Dissociative electron attachment to
  dna},}\ }\href@noop {} {\bibfield  {journal} {\bibinfo  {journal} {Phys. Rev.
  Lett.}\ }\textbf {\bibinfo {volume} {90}},\ \bibinfo {pages} {208102}
  (\bibinfo {year} {2003})}\BibitemShut {NoStop}%
\bibitem [{\citenamefont {Martin}\ \emph {et~al.}(2004)\citenamefont {Martin},
  \citenamefont {Burrow}, \citenamefont {Cai}, \citenamefont {Cloutier},
  \citenamefont {Hunting},\ and\ \citenamefont {Sanche}}]{Martin04}%
  \BibitemOpen
  \bibfield  {author} {\bibinfo {author} {\bibfnamefont {F.}~\bibnamefont
  {Martin}}, \bibinfo {author} {\bibfnamefont {P.~D.}\ \bibnamefont {Burrow}},
  \bibinfo {author} {\bibfnamefont {Z.}~\bibnamefont {Cai}}, \bibinfo {author}
  {\bibfnamefont {P.}~\bibnamefont {Cloutier}}, \bibinfo {author}
  {\bibfnamefont {D.}~\bibnamefont {Hunting}}, \ and\ \bibinfo {author}
  {\bibfnamefont {L.}~\bibnamefont {Sanche}},\ }\bibfield  {title} {\enquote
  {\bibinfo {title} {Dna strand breaks induced by 0--4 ev electrons: The role
  of shape resonances},}\ }\href@noop {} {\bibfield  {journal} {\bibinfo
  {journal} {Phys. Rev. Lett.}\ }\textbf {\bibinfo {volume} {93}},\ \bibinfo
  {pages} {068101} (\bibinfo {year} {2004})}\BibitemShut {NoStop}%
\bibitem [{\citenamefont {Surdutovich}\ and\ \citenamefont
  {Solov'yov}(2012)}]{Surdutovich12}%
  \BibitemOpen
  \bibfield  {author} {\bibinfo {author} {\bibfnamefont {E.}~\bibnamefont
  {Surdutovich}}\ and\ \bibinfo {author} {\bibfnamefont {A.~V.}\ \bibnamefont
  {Solov'yov}},\ }\bibfield  {title} {\enquote {\bibinfo {title} {{Double
  strand breaks in DNA resulting from double ionization events}},}\ }\href@noop
  {} {\bibfield  {journal} {\bibinfo  {journal} {Eur. Phys. J. D: Atomic,
  Molecular, Optical and Plasma Physics}\ }\textbf {\bibinfo {volume} {66}},\
  \bibinfo {pages} {206} (\bibinfo {year} {2012})}\BibitemShut {NoStop}%
\bibitem [{\citenamefont {Alizadeh}\ \emph {et~al.}(2015)\citenamefont
  {Alizadeh}, \citenamefont {Orlando},\ and\ \citenamefont
  {Sanche}}]{Alizadeh15}%
  \BibitemOpen
  \bibfield  {author} {\bibinfo {author} {\bibfnamefont {E.}~\bibnamefont
  {Alizadeh}}, \bibinfo {author} {\bibfnamefont {T.~M.}\ \bibnamefont
  {Orlando}}, \ and\ \bibinfo {author} {\bibfnamefont {L.}~\bibnamefont
  {Sanche}},\ }\bibfield  {title} {\enquote {\bibinfo {title} {{Biomolecular
  Damage Induced by Ionizing Radiation: The Direct and Indirect Effects of
  Low-Energy Electrons on DNA}},}\ }\href@noop {} {\bibfield  {journal}
  {\bibinfo  {journal} {Annu. Rev. Phys. Chem.}\ }\textbf {\bibinfo {volume}
  {66}},\ \bibinfo {pages} {379} (\bibinfo {year} {2015})}\BibitemShut
  {NoStop}%
\bibitem [{\citenamefont {Fasshauer}()}]{rapid_spec}%
  \BibitemOpen
  \bibfield  {author} {\bibinfo {author} {\bibfnamefont {E.}~\bibnamefont
  {Fasshauer}},\ }\href@noop {} {}\bibinfo {note}
  {DOI:10.6084/m9.figshare.7808294.v2}\BibitemShut {NoStop}%
\bibitem [{\citenamefont {Krausz}\ and\ \citenamefont
  {Ivanov}(2009)}]{Krausz09}%
  \BibitemOpen
  \bibfield  {author} {\bibinfo {author} {\bibfnamefont {Ferenc}\ \bibnamefont
  {Krausz}}\ and\ \bibinfo {author} {\bibfnamefont {Misha}\ \bibnamefont
  {Ivanov}},\ }\bibfield  {title} {\enquote {\bibinfo {title} {Attosecond
  physics},}\ }\href {\doibase 10.1103/RevModPhys.81.163} {\bibfield  {journal}
  {\bibinfo  {journal} {Rev. Mod. Phys.}\ }\textbf {\bibinfo {volume} {81}},\
  \bibinfo {pages} {163--234} (\bibinfo {year} {2009})}\BibitemShut {NoStop}%
\bibitem [{\citenamefont {Kaldun}\ \emph {et~al.}(2016)\citenamefont {Kaldun},
  \citenamefont {Bl{\"a}ttermann}, \citenamefont {Stoo\ss}, \citenamefont
  {Donsa}, \citenamefont {Wei}, \citenamefont {Pazourek}, \citenamefont
  {Nagele}, \citenamefont {Ott}, \citenamefont {Burgd{\"o}rfer},\ and\
  \citenamefont {Pfeifer}}]{Kaldun16}%
  \BibitemOpen
  \bibfield  {author} {\bibinfo {author} {\bibfnamefont {A.}~\bibnamefont
  {Kaldun}}, \bibinfo {author} {\bibfnamefont {A.}~\bibnamefont
  {Bl{\"a}ttermann}}, \bibinfo {author} {\bibfnamefont {V.}~\bibnamefont
  {Stoo\ss}}, \bibinfo {author} {\bibfnamefont {S.}~\bibnamefont {Donsa}},
  \bibinfo {author} {\bibfnamefont {H.}~\bibnamefont {Wei}}, \bibinfo {author}
  {\bibfnamefont {R.}~\bibnamefont {Pazourek}}, \bibinfo {author}
  {\bibfnamefont {S.}~\bibnamefont {Nagele}}, \bibinfo {author} {\bibfnamefont
  {C.~D.}\ \bibnamefont {Ott}, \bibfnamefont {C.~Lin}}, \bibinfo {author}
  {\bibfnamefont {J.}~\bibnamefont {Burgd{\"o}rfer}}, \ and\ \bibinfo {author}
  {\bibfnamefont {T.}~\bibnamefont {Pfeifer}},\ }\bibfield  {title} {\enquote
  {\bibinfo {title} {Observing the ultrafast buildup of a fano resonance in the
  time domain},}\ }\href@noop {} {\bibfield  {journal} {\bibinfo  {journal}
  {Science}\ }\textbf {\bibinfo {volume} {354}},\ \bibinfo {pages} {738}
  (\bibinfo {year} {2016})}\BibitemShut {NoStop}%
\bibitem [{\citenamefont {Gruson}\ \emph {et~al.}(2016)\citenamefont {Gruson},
  \citenamefont {Barreau}, \citenamefont {Jimenez-Galan}, \citenamefont
  {Risoud}, \citenamefont {Caillat}, \citenamefont {Maquet}, \citenamefont
  {Carre}, \citenamefont {Lepetit}, \citenamefont {Hergott}, \citenamefont
  {Ruchon}, \citenamefont {Argenti}, \citenamefont {Taieb}, \citenamefont
  {Martin},\ and\ \citenamefont {Salieres}}]{Gruson16}%
  \BibitemOpen
  \bibfield  {author} {\bibinfo {author} {\bibfnamefont {V.}~\bibnamefont
  {Gruson}}, \bibinfo {author} {\bibfnamefont {L.}~\bibnamefont {Barreau}},
  \bibinfo {author} {\bibfnamefont {A.}~\bibnamefont {Jimenez-Galan}}, \bibinfo
  {author} {\bibfnamefont {F.}~\bibnamefont {Risoud}}, \bibinfo {author}
  {\bibfnamefont {J.}~\bibnamefont {Caillat}}, \bibinfo {author} {\bibfnamefont
  {A.}~\bibnamefont {Maquet}}, \bibinfo {author} {\bibfnamefont
  {B.}~\bibnamefont {Carre}}, \bibinfo {author} {\bibfnamefont
  {F.}~\bibnamefont {Lepetit}}, \bibinfo {author} {\bibfnamefont {J.-F.}\
  \bibnamefont {Hergott}}, \bibinfo {author} {\bibfnamefont {T.}~\bibnamefont
  {Ruchon}}, \bibinfo {author} {\bibfnamefont {L.}~\bibnamefont {Argenti}},
  \bibinfo {author} {\bibfnamefont {R.}~\bibnamefont {Taieb}}, \bibinfo
  {author} {\bibfnamefont {F.}~\bibnamefont {Martin}}, \ and\ \bibinfo {author}
  {\bibfnamefont {P.}~\bibnamefont {Salieres}},\ }\bibfield  {title} {\enquote
  {\bibinfo {title} {Attosecond dynamics through a fano resonance: Monitoring
  the birth of a photoelectron},}\ }\href@noop {} {\bibfield  {journal}
  {\bibinfo  {journal} {Science}\ }\textbf {\bibinfo {volume} {354}},\ \bibinfo
  {pages} {734} (\bibinfo {year} {2016})}\BibitemShut {NoStop}%
\bibitem [{\citenamefont {Kuleff}\ and\ \citenamefont
  {Cederbaum}(2007)}]{Kuleff07}%
  \BibitemOpen
  \bibfield  {author} {\bibinfo {author} {\bibfnamefont {A.~I.}\ \bibnamefont
  {Kuleff}}\ and\ \bibinfo {author} {\bibfnamefont {L.~S.}\ \bibnamefont
  {Cederbaum}},\ }\bibfield  {title} {\enquote {\bibinfo {title} {{Tracing
  Ultrafast Interatomic Electronic Decay Processes in Real Time and Space}},}\
  }\href {\doibase http://dx.doi.org/10.1103/PhysRevLett.98.083201} {\bibfield
  {journal} {\bibinfo  {journal} {Phys. Rev. Lett.}\ }\textbf {\bibinfo
  {volume} {98}},\ \bibinfo {pages} {083201} (\bibinfo {year}
  {2007})}\BibitemShut {NoStop}%
\bibitem [{\citenamefont {Schnorr}\ \emph {et~al.}(2013)\citenamefont
  {Schnorr}, \citenamefont {Senftleben}, \citenamefont {Kurka}, \citenamefont
  {Rudenko}, \citenamefont {Foucar}, \citenamefont {Schmid}, \citenamefont
  {Broska}, \citenamefont {Pfeifer}, \citenamefont {Meyer}, \citenamefont
  {Anielski}, \citenamefont {Boll}, \citenamefont {Rolles}, \citenamefont
  {K{\"u}bel}, \citenamefont {Kling}, \citenamefont {Jiang}, \citenamefont
  {Mondal}, \citenamefont {Tachibana}, \citenamefont {Ueda}, \citenamefont
  {Marchenko}, \citenamefont {Simon}, \citenamefont {Brenner}, \citenamefont
  {Treusch}, \citenamefont {Scheit}, \citenamefont {Averbukh}, \citenamefont
  {Ullrich}, \citenamefont {Schr{\"o}ter},\ and\ \citenamefont
  {Moshammer}}]{Schnorr13}%
  \BibitemOpen
  \bibfield  {author} {\bibinfo {author} {\bibfnamefont {K.}~\bibnamefont
  {Schnorr}}, \bibinfo {author} {\bibfnamefont {A.}~\bibnamefont {Senftleben}},
  \bibinfo {author} {\bibfnamefont {M.}~\bibnamefont {Kurka}}, \bibinfo
  {author} {\bibfnamefont {A.}~\bibnamefont {Rudenko}}, \bibinfo {author}
  {\bibfnamefont {L.}~\bibnamefont {Foucar}}, \bibinfo {author} {\bibfnamefont
  {G.}~\bibnamefont {Schmid}}, \bibinfo {author} {\bibfnamefont
  {A.}~\bibnamefont {Broska}}, \bibinfo {author} {\bibfnamefont
  {T.}~\bibnamefont {Pfeifer}}, \bibinfo {author} {\bibfnamefont
  {K.}~\bibnamefont {Meyer}}, \bibinfo {author} {\bibfnamefont
  {D.}~\bibnamefont {Anielski}}, \bibinfo {author} {\bibfnamefont
  {R.}~\bibnamefont {Boll}}, \bibinfo {author} {\bibfnamefont {D.}~\bibnamefont
  {Rolles}}, \bibinfo {author} {\bibfnamefont {M.}~\bibnamefont {K{\"u}bel}},
  \bibinfo {author} {\bibfnamefont {M.~F.}\ \bibnamefont {Kling}}, \bibinfo
  {author} {\bibfnamefont {Y.~H.}\ \bibnamefont {Jiang}}, \bibinfo {author}
  {\bibfnamefont {S.}~\bibnamefont {Mondal}}, \bibinfo {author} {\bibfnamefont
  {T.}~\bibnamefont {Tachibana}}, \bibinfo {author} {\bibfnamefont
  {K.}~\bibnamefont {Ueda}}, \bibinfo {author} {\bibfnamefont {T.}~\bibnamefont
  {Marchenko}}, \bibinfo {author} {\bibfnamefont {M.}~\bibnamefont {Simon}},
  \bibinfo {author} {\bibfnamefont {G.}~\bibnamefont {Brenner}}, \bibinfo
  {author} {\bibfnamefont {R.}~\bibnamefont {Treusch}}, \bibinfo {author}
  {\bibfnamefont {S.}~\bibnamefont {Scheit}}, \bibinfo {author} {\bibfnamefont
  {V.}~\bibnamefont {Averbukh}}, \bibinfo {author} {\bibfnamefont
  {J.}~\bibnamefont {Ullrich}}, \bibinfo {author} {\bibfnamefont {C.~D.}\
  \bibnamefont {Schr{\"o}ter}}, \ and\ \bibinfo {author} {\bibfnamefont
  {R.}~\bibnamefont {Moshammer}},\ }\bibfield  {title} {\enquote {\bibinfo
  {title} {{Time-Resolved Measurement of Interatomic Coulombic Decay in
  Ne$_2$}},}\ }\href@noop {} {\bibfield  {journal} {\bibinfo  {journal} {Phys.
  Rev. Lett.}\ }\textbf {\bibinfo {volume} {111}},\ \bibinfo {pages} {093402}
  (\bibinfo {year} {2013})}\BibitemShut {NoStop}%
\bibitem [{\citenamefont {Trinter}\ \emph {et~al.}(2013)\citenamefont
  {Trinter}, \citenamefont {Williams}, \citenamefont {Weller}, \citenamefont
  {Waitz}, \citenamefont {Pitzer}, \citenamefont {Voigtsberger}, \citenamefont
  {Schober}, \citenamefont {Kastirke}, \citenamefont {M{\"u}ller},
  \citenamefont {Goihl}, \citenamefont {Burzynski}, \citenamefont {Wiegandt},
  \citenamefont {Bauer}, \citenamefont {Wallauer}, \citenamefont {Sann},
  \citenamefont {Kalinin}, \citenamefont {Schmidt}, \citenamefont
  {Sch{\"o}ffler}, \citenamefont {Sisourat},\ and\ \citenamefont
  {Jahnke}}]{Trinter13a}%
  \BibitemOpen
  \bibfield  {author} {\bibinfo {author} {\bibfnamefont {F.}~\bibnamefont
  {Trinter}}, \bibinfo {author} {\bibfnamefont {J.~B.}\ \bibnamefont
  {Williams}}, \bibinfo {author} {\bibfnamefont {M.}~\bibnamefont {Weller}},
  \bibinfo {author} {\bibfnamefont {M.}~\bibnamefont {Waitz}}, \bibinfo
  {author} {\bibfnamefont {M.}~\bibnamefont {Pitzer}}, \bibinfo {author}
  {\bibfnamefont {J.}~\bibnamefont {Voigtsberger}}, \bibinfo {author}
  {\bibfnamefont {C.}~\bibnamefont {Schober}}, \bibinfo {author} {\bibfnamefont
  {G.}~\bibnamefont {Kastirke}}, \bibinfo {author} {\bibfnamefont
  {C.}~\bibnamefont {M{\"u}ller}}, \bibinfo {author} {\bibfnamefont
  {C.}~\bibnamefont {Goihl}}, \bibinfo {author} {\bibfnamefont
  {P.}~\bibnamefont {Burzynski}}, \bibinfo {author} {\bibfnamefont
  {F.}~\bibnamefont {Wiegandt}}, \bibinfo {author} {\bibfnamefont
  {T.}~\bibnamefont {Bauer}}, \bibinfo {author} {\bibfnamefont
  {R.}~\bibnamefont {Wallauer}}, \bibinfo {author} {\bibfnamefont
  {H.}~\bibnamefont {Sann}}, \bibinfo {author} {\bibfnamefont {A.}~\bibnamefont
  {Kalinin}}, \bibinfo {author} {\bibfnamefont {L.~Ph.~H.}\ \bibnamefont
  {Schmidt}}, \bibinfo {author} {\bibfnamefont {M.}~\bibnamefont
  {Sch{\"o}ffler}}, \bibinfo {author} {\bibfnamefont {N.}~\bibnamefont
  {Sisourat}}, \ and\ \bibinfo {author} {\bibfnamefont {T.}~\bibnamefont
  {Jahnke}},\ }\bibfield  {title} {\enquote {\bibinfo {title} {{Evolution of
  Interatomic Coulombic Decay in the Time Domain}},}\ }\href {\doibase
  http://dx.doi.org/10.1103/PhysRevLett.111.093401} {\bibfield  {journal}
  {\bibinfo  {journal} {Phys. Rev. Lett.}\ }\textbf {\bibinfo {volume} {111}},\
  \bibinfo {pages} {093401} (\bibinfo {year} {2013})}\BibitemShut {NoStop}%
\bibitem [{\citenamefont {Schnorr}\ \emph {et~al.}(2015)\citenamefont
  {Schnorr}, \citenamefont {Senftleben}, \citenamefont {Schmid}, \citenamefont
  {Augustin}, \citenamefont {Kurka}, \citenamefont {Rudenko}, \citenamefont
  {Foucar}, \citenamefont {Broska}, \citenamefont {Meyer}, \citenamefont
  {Anielski}, \citenamefont {Anielski}, \citenamefont {Boll}, \citenamefont
  {Rolles}, \citenamefont {K{\"u}bel}, \citenamefont {Kling}, \citenamefont
  {Jiang}, \citenamefont {Mondal}, \citenamefont {Tachibana}, \citenamefont
  {Ueda}, \citenamefont {Marchenko}, \citenamefont {Simon}, \citenamefont
  {Brenner}, \citenamefont {Treusch}, \citenamefont {Scheit}, \citenamefont
  {Averbukh}, \citenamefont {Ullrich}, \citenamefont {Pfeifer}, \citenamefont
  {Schr{\"o}ter},\ and\ \citenamefont {Moshammer}}]{Schnorr15}%
  \BibitemOpen
  \bibfield  {author} {\bibinfo {author} {\bibfnamefont {K.}~\bibnamefont
  {Schnorr}}, \bibinfo {author} {\bibfnamefont {A.}~\bibnamefont {Senftleben}},
  \bibinfo {author} {\bibfnamefont {G.}~\bibnamefont {Schmid}}, \bibinfo
  {author} {\bibfnamefont {S.}~\bibnamefont {Augustin}}, \bibinfo {author}
  {\bibfnamefont {M.}~\bibnamefont {Kurka}}, \bibinfo {author} {\bibfnamefont
  {A.}~\bibnamefont {Rudenko}}, \bibinfo {author} {\bibfnamefont
  {L.}~\bibnamefont {Foucar}}, \bibinfo {author} {\bibfnamefont
  {A.}~\bibnamefont {Broska}}, \bibinfo {author} {\bibfnamefont
  {K.}~\bibnamefont {Meyer}}, \bibinfo {author} {\bibfnamefont
  {D.}~\bibnamefont {Anielski}}, \bibinfo {author} {\bibfnamefont
  {D.}~\bibnamefont {Anielski}}, \bibinfo {author} {\bibfnamefont
  {R.}~\bibnamefont {Boll}}, \bibinfo {author} {\bibfnamefont {D.}~\bibnamefont
  {Rolles}}, \bibinfo {author} {\bibfnamefont {M.}~\bibnamefont {K{\"u}bel}},
  \bibinfo {author} {\bibfnamefont {M.~F.}\ \bibnamefont {Kling}}, \bibinfo
  {author} {\bibfnamefont {Y.~H.}\ \bibnamefont {Jiang}}, \bibinfo {author}
  {\bibfnamefont {S.}~\bibnamefont {Mondal}}, \bibinfo {author} {\bibfnamefont
  {T.}~\bibnamefont {Tachibana}}, \bibinfo {author} {\bibfnamefont
  {K.}~\bibnamefont {Ueda}}, \bibinfo {author} {\bibfnamefont {T.}~\bibnamefont
  {Marchenko}}, \bibinfo {author} {\bibfnamefont {M.}~\bibnamefont {Simon}},
  \bibinfo {author} {\bibfnamefont {G.}~\bibnamefont {Brenner}}, \bibinfo
  {author} {\bibfnamefont {R.}~\bibnamefont {Treusch}}, \bibinfo {author}
  {\bibfnamefont {S.}~\bibnamefont {Scheit}}, \bibinfo {author} {\bibfnamefont
  {V.}~\bibnamefont {Averbukh}}, \bibinfo {author} {\bibfnamefont
  {J.}~\bibnamefont {Ullrich}}, \bibinfo {author} {\bibfnamefont
  {T.}~\bibnamefont {Pfeifer}}, \bibinfo {author} {\bibfnamefont {C.~D.}\
  \bibnamefont {Schr{\"o}ter}}, \ and\ \bibinfo {author} {\bibfnamefont
  {R.}~\bibnamefont {Moshammer}},\ }\bibfield  {title} {\enquote {\bibinfo
  {title} {{Time-resolved study of ICD in Ne dimers using FEL radiation}},}\
  }\href@noop {} {\bibfield  {journal} {\bibinfo  {journal} {J. Electron
  Spectrosc. Relat. Phenom.}\ }\textbf {\bibinfo {volume} {204}},\ \bibinfo
  {pages} {245} (\bibinfo {year} {2015})}\BibitemShut {NoStop}%
\bibitem [{\citenamefont {Frühling}\ \emph {et~al.}(2015)\citenamefont
  {Frühling}, \citenamefont {Trinter}, \citenamefont {Karimi}, \citenamefont
  {Williams},\ and\ \citenamefont {Jahnke}}]{Fruehling15}%
  \BibitemOpen
  \bibfield  {author} {\bibinfo {author} {\bibfnamefont {U.}~\bibnamefont
  {Frühling}}, \bibinfo {author} {\bibfnamefont {F.}~\bibnamefont {Trinter}},
  \bibinfo {author} {\bibfnamefont {F.}~\bibnamefont {Karimi}}, \bibinfo
  {author} {\bibfnamefont {J.~B.}\ \bibnamefont {Williams}}, \ and\ \bibinfo
  {author} {\bibfnamefont {T.}~\bibnamefont {Jahnke}},\ }\bibfield  {title}
  {\enquote {\bibinfo {title} {{Time-resolved studies of Interatomic Coulombic
  Decay}},}\ }\href {\doibase http://dx.doi.org/10.1016/j.elspec.2015.06.012}
  {\bibfield  {journal} {\bibinfo  {journal} {J. Electron. Spectrosc. Relat.
  Phenom.}\ }\textbf {\bibinfo {volume} {204}},\ \bibinfo {pages} {237}
  (\bibinfo {year} {2015})}\BibitemShut {NoStop}%
\bibitem [{\citenamefont {Mizuno}\ \emph {et~al.}(2017)\citenamefont {Mizuno},
  \citenamefont {C{\"o}rlin}, \citenamefont {Miteva}, \citenamefont {Gokhberg},
  \citenamefont {Kuleff}, \citenamefont {Cederbaum}, \citenamefont {Pfeifer},
  \citenamefont {Fischer},\ and\ \citenamefont {Moshammer}}]{Mizuno17}%
  \BibitemOpen
  \bibfield  {author} {\bibinfo {author} {\bibfnamefont {T.}~\bibnamefont
  {Mizuno}}, \bibinfo {author} {\bibfnamefont {P.}~\bibnamefont {C{\"o}rlin}},
  \bibinfo {author} {\bibfnamefont {T.}~\bibnamefont {Miteva}}, \bibinfo
  {author} {\bibfnamefont {K.}~\bibnamefont {Gokhberg}}, \bibinfo {author}
  {\bibfnamefont {A.}~\bibnamefont {Kuleff}}, \bibinfo {author} {\bibfnamefont
  {L.~S.}\ \bibnamefont {Cederbaum}}, \bibinfo {author} {\bibfnamefont
  {T.}~\bibnamefont {Pfeifer}}, \bibinfo {author} {\bibfnamefont
  {A.}~\bibnamefont {Fischer}}, \ and\ \bibinfo {author} {\bibfnamefont
  {R.}~\bibnamefont {Moshammer}},\ }\bibfield  {title} {\enquote {\bibinfo
  {title} {Time-resolved observation of interatomic excitation-energy transfer
  in argon dimers},}\ }\href@noop {} {\bibfield  {journal} {\bibinfo  {journal}
  {J. Chem. Phys.}\ }\textbf {\bibinfo {volume} {146}},\ \bibinfo {pages}
  {104305} (\bibinfo {year} {2017})}\BibitemShut {NoStop}%
\bibitem [{\citenamefont {Takanashi}\ \emph {et~al.}(2017)\citenamefont
  {Takanashi}, \citenamefont {Golubev}, \citenamefont {Callegari},
  \citenamefont {Fukuzawa}, \citenamefont {Motomura}, \citenamefont
  {Iablonskyi}, \citenamefont {Kumagai}, \citenamefont {Mondal}, \citenamefont
  {Tachibana}, \citenamefont {Nagaya}, \citenamefont {Nishiyama}, \citenamefont
  {Matsunami}, \citenamefont {Johnsson}, \citenamefont {Piseri}, \citenamefont
  {Sansone}, \citenamefont {Dubrouil}, \citenamefont {Reduzzi}, \citenamefont
  {Carpeggiani}, \citenamefont {Vozzi}, \citenamefont {Devetta}, \citenamefont
  {Negro}, \citenamefont {Facciala}, \citenamefont {Calegari}, \citenamefont
  {Trabattoni}, \citenamefont {Castrovilli}, \citenamefont {Ovcharenko},
  \citenamefont {Mudrich}, \citenamefont {Stienkemeier}, \citenamefont
  {Coreno}, \citenamefont {Alagia}, \citenamefont {Sch{\"u}tte}, \citenamefont
  {Berrah}, \citenamefont {Plekan}, \citenamefont {Finetti}, \citenamefont
  {Spezzani}, \citenamefont {Ferrari}, \citenamefont {Allaria}, \citenamefont
  {Penco}, \citenamefont {Serpico}, \citenamefont {De~Ninno}, \citenamefont
  {Diviacco}, \citenamefont {Di~Mitri}, \citenamefont {Giannessi},
  \citenamefont {Jabbari}, \citenamefont {Prince}, \citenamefont {Cederbaum},
  \citenamefont {Demekhin}, \citenamefont {Kuleff},\ and\ \citenamefont
  {Ueda}}]{Takanashi17}%
  \BibitemOpen
  \bibfield  {author} {\bibinfo {author} {\bibfnamefont {T.}~\bibnamefont
  {Takanashi}}, \bibinfo {author} {\bibfnamefont {N.~V.}\ \bibnamefont
  {Golubev}}, \bibinfo {author} {\bibfnamefont {C.}~\bibnamefont {Callegari}},
  \bibinfo {author} {\bibfnamefont {H.}~\bibnamefont {Fukuzawa}}, \bibinfo
  {author} {\bibfnamefont {K.}~\bibnamefont {Motomura}}, \bibinfo {author}
  {\bibfnamefont {D.}~\bibnamefont {Iablonskyi}}, \bibinfo {author}
  {\bibfnamefont {Y.}~\bibnamefont {Kumagai}}, \bibinfo {author} {\bibfnamefont
  {S.}~\bibnamefont {Mondal}}, \bibinfo {author} {\bibfnamefont
  {T.}~\bibnamefont {Tachibana}}, \bibinfo {author} {\bibfnamefont
  {K.}~\bibnamefont {Nagaya}}, \bibinfo {author} {\bibfnamefont
  {T.}~\bibnamefont {Nishiyama}}, \bibinfo {author} {\bibfnamefont
  {K.}~\bibnamefont {Matsunami}}, \bibinfo {author} {\bibfnamefont
  {P.}~\bibnamefont {Johnsson}}, \bibinfo {author} {\bibfnamefont
  {P.}~\bibnamefont {Piseri}}, \bibinfo {author} {\bibfnamefont
  {G.}~\bibnamefont {Sansone}}, \bibinfo {author} {\bibfnamefont
  {A.}~\bibnamefont {Dubrouil}}, \bibinfo {author} {\bibfnamefont
  {M.}~\bibnamefont {Reduzzi}}, \bibinfo {author} {\bibfnamefont
  {P.}~\bibnamefont {Carpeggiani}}, \bibinfo {author} {\bibfnamefont
  {C.}~\bibnamefont {Vozzi}}, \bibinfo {author} {\bibfnamefont
  {M.}~\bibnamefont {Devetta}}, \bibinfo {author} {\bibfnamefont
  {M.}~\bibnamefont {Negro}}, \bibinfo {author} {\bibfnamefont
  {D.}~\bibnamefont {Facciala}}, \bibinfo {author} {\bibfnamefont
  {F.}~\bibnamefont {Calegari}}, \bibinfo {author} {\bibfnamefont
  {A.}~\bibnamefont {Trabattoni}}, \bibinfo {author} {\bibfnamefont {M.~C.}\
  \bibnamefont {Castrovilli}}, \bibinfo {author} {\bibfnamefont
  {Y.}~\bibnamefont {Ovcharenko}}, \bibinfo {author} {\bibfnamefont
  {M.}~\bibnamefont {Mudrich}}, \bibinfo {author} {\bibfnamefont
  {F.}~\bibnamefont {Stienkemeier}}, \bibinfo {author} {\bibfnamefont
  {M.}~\bibnamefont {Coreno}}, \bibinfo {author} {\bibfnamefont
  {M.}~\bibnamefont {Alagia}}, \bibinfo {author} {\bibfnamefont
  {B.}~\bibnamefont {Sch{\"u}tte}}, \bibinfo {author} {\bibfnamefont
  {N.}~\bibnamefont {Berrah}}, \bibinfo {author} {\bibfnamefont
  {O.}~\bibnamefont {Plekan}}, \bibinfo {author} {\bibfnamefont
  {P.}~\bibnamefont {Finetti}}, \bibinfo {author} {\bibfnamefont
  {C.}~\bibnamefont {Spezzani}}, \bibinfo {author} {\bibfnamefont
  {E.}~\bibnamefont {Ferrari}}, \bibinfo {author} {\bibfnamefont
  {E.}~\bibnamefont {Allaria}}, \bibinfo {author} {\bibfnamefont
  {G.}~\bibnamefont {Penco}}, \bibinfo {author} {\bibfnamefont
  {C.}~\bibnamefont {Serpico}}, \bibinfo {author} {\bibfnamefont
  {G.}~\bibnamefont {De~Ninno}}, \bibinfo {author} {\bibfnamefont
  {B.}~\bibnamefont {Diviacco}}, \bibinfo {author} {\bibfnamefont
  {S.}~\bibnamefont {Di~Mitri}}, \bibinfo {author} {\bibfnamefont
  {L.}~\bibnamefont {Giannessi}}, \bibinfo {author} {\bibfnamefont
  {G.}~\bibnamefont {Jabbari}}, \bibinfo {author} {\bibfnamefont {K.~C.}\
  \bibnamefont {Prince}}, \bibinfo {author} {\bibfnamefont {L.~S.}\
  \bibnamefont {Cederbaum}}, \bibinfo {author} {\bibfnamefont {Ph.~V.}\
  \bibnamefont {Demekhin}}, \bibinfo {author} {\bibfnamefont {A.~I.}\
  \bibnamefont {Kuleff}}, \ and\ \bibinfo {author} {\bibfnamefont
  {K.}~\bibnamefont {Ueda}},\ }\bibfield  {title} {\enquote {\bibinfo {title}
  {{Time-Resolved Measurement of Interatomic Coulombic Decay Induced by
  Two-Photon Double Excitation of Ne2}},}\ }\href {\doibase
  http://dx.doi.org/10.1103/PhysRevLett.118.033202} {\bibfield  {journal}
  {\bibinfo  {journal} {Phys. Rev. Lett.}\ }\textbf {\bibinfo {volume} {118}},\
  \bibinfo {pages} {033202} (\bibinfo {year} {2017})}\BibitemShut {NoStop}%
\bibitem [{\citenamefont {Jing}\ and\ \citenamefont {Madsen}(2019)}]{Qingli19}%
  \BibitemOpen
  \bibfield  {author} {\bibinfo {author} {\bibfnamefont {Qingli}\ \bibnamefont
  {Jing}}\ and\ \bibinfo {author} {\bibfnamefont {Lars~Bojer}\ \bibnamefont
  {Madsen}},\ }\bibfield  {title} {\enquote {\bibinfo {title} {Dynamics of
  interatomic coulombic decay in neon dimers by xuv-pump--xuv-probe
  spectroscopy},}\ }\href {\doibase 10.1103/PhysRevA.99.013409} {\bibfield
  {journal} {\bibinfo  {journal} {Phys. Rev. A}\ }\textbf {\bibinfo {volume}
  {99}},\ \bibinfo {pages} {013409} (\bibinfo {year} {2019})}\BibitemShut
  {NoStop}%
\bibitem [{\citenamefont {Gokhberg}\ \emph {et~al.}(2006)\citenamefont
  {Gokhberg}, \citenamefont {Averbukh},\ and\ \citenamefont
  {Cederbaum}}]{Gokhberg06}%
  \BibitemOpen
  \bibfield  {author} {\bibinfo {author} {\bibfnamefont {K.}~\bibnamefont
  {Gokhberg}}, \bibinfo {author} {\bibfnamefont {V.}~\bibnamefont {Averbukh}},
  \ and\ \bibinfo {author} {\bibfnamefont {L.~S.}\ \bibnamefont {Cederbaum}},\
  }\bibfield  {title} {\enquote {\bibinfo {title} {{Interatomic decay of
  inner-valence-excited states in clusters}},}\ }\href@noop {} {\bibfield
  {journal} {\bibinfo  {journal} {J. Chem. Phys.}\ }\textbf {\bibinfo {volume}
  {124}},\ \bibinfo {pages} {144315} (\bibinfo {year} {2006})}\BibitemShut
  {NoStop}%
\bibitem [{\citenamefont {Barth}\ \emph {et~al.}(2005)\citenamefont {Barth},
  \citenamefont {Joshi}, \citenamefont {Marburger}, \citenamefont {Ulrich},
  \citenamefont {Lindblad}, \citenamefont {{\"O}hrwall}, \citenamefont
  {Bj{\"o}rneholm},\ and\ \citenamefont {Hergenhahn}}]{Barth05}%
  \BibitemOpen
  \bibfield  {author} {\bibinfo {author} {\bibfnamefont {S.}~\bibnamefont
  {Barth}}, \bibinfo {author} {\bibfnamefont {S.}~\bibnamefont {Joshi}},
  \bibinfo {author} {\bibfnamefont {S.}~\bibnamefont {Marburger}}, \bibinfo
  {author} {\bibfnamefont {V.}~\bibnamefont {Ulrich}}, \bibinfo {author}
  {\bibfnamefont {A.}~\bibnamefont {Lindblad}}, \bibinfo {author}
  {\bibfnamefont {G.}~\bibnamefont {{\"O}hrwall}}, \bibinfo {author}
  {\bibfnamefont {O.}~\bibnamefont {Bj{\"o}rneholm}}, \ and\ \bibinfo {author}
  {\bibfnamefont {U.}~\bibnamefont {Hergenhahn}},\ }\bibfield  {title}
  {\enquote {\bibinfo {title} {{Observation of resonant Interatomic Coulombic
  Decay in Ne clusters}},}\ }\href {\doibase
  http://dx.doi.org/10.1063/1.1937395} {\bibfield  {journal} {\bibinfo
  {journal} {J. Chem. Phys.}\ }\textbf {\bibinfo {volume} {122}},\ \bibinfo
  {eid} {4} (\bibinfo {year} {2005})}\BibitemShut {NoStop}%
\bibitem [{\citenamefont {Aoto}\ \emph {et~al.}(2006)\citenamefont {Aoto},
  \citenamefont {Ito}, \citenamefont {Hikosaka}, \citenamefont {Shigemasa},
  \citenamefont {Penent},\ and\ \citenamefont {Lablanquie}}]{aoto2006}%
  \BibitemOpen
  \bibfield  {author} {\bibinfo {author} {\bibfnamefont {T.}~\bibnamefont
  {Aoto}}, \bibinfo {author} {\bibfnamefont {K.}~\bibnamefont {Ito}}, \bibinfo
  {author} {\bibfnamefont {Y.}~\bibnamefont {Hikosaka}}, \bibinfo {author}
  {\bibfnamefont {E.}~\bibnamefont {Shigemasa}}, \bibinfo {author}
  {\bibfnamefont {F.}~\bibnamefont {Penent}}, \ and\ \bibinfo {author}
  {\bibfnamefont {P.}~\bibnamefont {Lablanquie}},\ }\bibfield  {title}
  {\enquote {\bibinfo {title} {Properties of resonant interatomic coulombic
  decay in ne dimers},}\ }\href@noop {} {\bibfield  {journal} {\bibinfo
  {journal} {Phys. Rev. Lett.}\ }\textbf {\bibinfo {volume} {97}},\ \bibinfo
  {pages} {243401} (\bibinfo {year} {2006})}\BibitemShut {NoStop}%
\bibitem [{\citenamefont {Kopelke}\ \emph {et~al.}(2009)\citenamefont
  {Kopelke}, \citenamefont {Gokhberg}, \citenamefont {Cederbaum},\ and\
  \citenamefont {Averbukh}}]{Kopelke09}%
  \BibitemOpen
  \bibfield  {author} {\bibinfo {author} {\bibfnamefont {S.}~\bibnamefont
  {Kopelke}}, \bibinfo {author} {\bibfnamefont {K.}~\bibnamefont {Gokhberg}},
  \bibinfo {author} {\bibfnamefont {L.~S.}\ \bibnamefont {Cederbaum}}, \ and\
  \bibinfo {author} {\bibfnamefont {V.}~\bibnamefont {Averbukh}},\ }\bibfield
  {title} {\enquote {\bibinfo {title} {Calculation of resonant interatomic
  coulombic decay widths of inner-valence-excited states delocalized due to
  inversion symmetry},}\ }\href@noop {} {\bibfield  {journal} {\bibinfo
  {journal} {J. Chem. Phys.}\ }\textbf {\bibinfo {volume} {130}},\ \bibinfo
  {pages} {144103} (\bibinfo {year} {2009})}\BibitemShut {NoStop}%
\bibitem [{\citenamefont {Knie}\ \emph {et~al.}(2014)\citenamefont {Knie},
  \citenamefont {Hans}, \citenamefont {Förstel}, \citenamefont {Hergenhahn},
  \citenamefont {Schmidt}, \citenamefont {Ph.}, \citenamefont {Ozga},
  \citenamefont {Kambs}, \citenamefont {Trinter}, \citenamefont {Voigtsberger},
  \citenamefont {Metz}, \citenamefont {Jahnke}, \citenamefont {Dörner},
  \citenamefont {Kuleff}, \citenamefont {Cederbaum}, \citenamefont {Demekhin},\
  and\ \citenamefont {Ehresmann}}]{Knie14}%
  \BibitemOpen
  \bibfield  {author} {\bibinfo {author} {\bibfnamefont {A.}~\bibnamefont
  {Knie}}, \bibinfo {author} {\bibfnamefont {A.}~\bibnamefont {Hans}}, \bibinfo
  {author} {\bibfnamefont {M.}~\bibnamefont {Förstel}}, \bibinfo {author}
  {\bibfnamefont {U.}~\bibnamefont {Hergenhahn}}, \bibinfo {author}
  {\bibfnamefont {P.}~\bibnamefont {Schmidt}}, \bibinfo {author} {\bibfnamefont
  {Reiß}\ \bibnamefont {Ph.}}, \bibinfo {author} {\bibfnamefont {Ch.}\
  \bibnamefont {Ozga}}, \bibinfo {author} {\bibfnamefont {B.}~\bibnamefont
  {Kambs}}, \bibinfo {author} {\bibfnamefont {F.}~\bibnamefont {Trinter}},
  \bibinfo {author} {\bibfnamefont {J.}~\bibnamefont {Voigtsberger}}, \bibinfo
  {author} {\bibfnamefont {D.}~\bibnamefont {Metz}}, \bibinfo {author}
  {\bibfnamefont {T.}~\bibnamefont {Jahnke}}, \bibinfo {author} {\bibfnamefont
  {R.}~\bibnamefont {Dörner}}, \bibinfo {author} {\bibfnamefont {A.~I.}\
  \bibnamefont {Kuleff}}, \bibinfo {author} {\bibfnamefont {L.~S.}\
  \bibnamefont {Cederbaum}}, \bibinfo {author} {\bibfnamefont {P.~V.}\
  \bibnamefont {Demekhin}}, \ and\ \bibinfo {author} {\bibfnamefont
  {A.}~\bibnamefont {Ehresmann}},\ }\bibfield  {title} {\enquote {\bibinfo
  {title} {{Detecting ultrafast interatomic electronic processes in media by
  fluorescence}},}\ }\href {\doibase
  http://dx.doi.org/10.1088/1367-2630/16/10/102002} {\bibfield  {journal}
  {\bibinfo  {journal} {New. J. Phys.}\ }\textbf {\bibinfo {volume} {16}},\
  \bibinfo {pages} {102002} (\bibinfo {year} {2014})}\BibitemShut {NoStop}%
\bibitem [{\citenamefont {Hans}\ \emph {et~al.}(2017)\citenamefont {Hans},
  \citenamefont {Ltaief}, \citenamefont {Förstel}, \citenamefont {Schmidt},
  \citenamefont {Ozga}, \citenamefont {Reiß}, \citenamefont {Holzapfel},
  \citenamefont {Küstner-Wetekam}, \citenamefont {Wiegandt}, \citenamefont
  {Trinter}, \citenamefont {Hergenhahn}, \citenamefont {Jahnke}, \citenamefont
  {Dörner}, \citenamefont {Ehresmann}, \citenamefont {Demekhin},\ and\
  \citenamefont {Knie}}]{Hans17}%
  \BibitemOpen
  \bibfield  {author} {\bibinfo {author} {\bibfnamefont {A.}~\bibnamefont
  {Hans}}, \bibinfo {author} {\bibfnamefont {L.~B.}\ \bibnamefont {Ltaief}},
  \bibinfo {author} {\bibfnamefont {M.}~\bibnamefont {Förstel}}, \bibinfo
  {author} {\bibfnamefont {Ph.}\ \bibnamefont {Schmidt}}, \bibinfo {author}
  {\bibfnamefont {Ch.}\ \bibnamefont {Ozga}}, \bibinfo {author} {\bibfnamefont
  {Ph.}\ \bibnamefont {Reiß}}, \bibinfo {author} {\bibfnamefont
  {X.}~\bibnamefont {Holzapfel}}, \bibinfo {author} {\bibfnamefont
  {C.}~\bibnamefont {Küstner-Wetekam}}, \bibinfo {author} {\bibfnamefont
  {F.}~\bibnamefont {Wiegandt}}, \bibinfo {author} {\bibfnamefont
  {F.}~\bibnamefont {Trinter}}, \bibinfo {author} {\bibfnamefont
  {U.}~\bibnamefont {Hergenhahn}}, \bibinfo {author} {\bibfnamefont
  {T.}~\bibnamefont {Jahnke}}, \bibinfo {author} {\bibfnamefont
  {R.}~\bibnamefont {Dörner}}, \bibinfo {author} {\bibfnamefont
  {A.}~\bibnamefont {Ehresmann}}, \bibinfo {author} {\bibfnamefont {Ph.~V.}\
  \bibnamefont {Demekhin}}, \ and\ \bibinfo {author} {\bibfnamefont
  {A.}~\bibnamefont {Knie}},\ }\bibfield  {title} {\enquote {\bibinfo {title}
  {{Fluorescence cascades evoked by resonant interatomic Coulombic decay of
  inner-valence excited neon clusters}},}\ }\href {\doibase
  http://dx.doi.org/10.1016/j.chemphys.2016.06.016} {\bibfield  {journal}
  {\bibinfo  {journal} {Chem. Phys.}\ }\textbf {\bibinfo {volume} {482}},\
  \bibinfo {pages} {165} (\bibinfo {year} {2017})}\BibitemShut {NoStop}%
\bibitem [{\citenamefont {Fano}(1961)}]{Fano61}%
  \BibitemOpen
  \bibfield  {author} {\bibinfo {author} {\bibfnamefont {U.}~\bibnamefont
  {Fano}},\ }\bibfield  {title} {\enquote {\bibinfo {title} {Effects of
  configuration interaction on intensities and phase shifts},}\ }\href@noop {}
  {\bibfield  {journal} {\bibinfo  {journal} {Phys. Rev.}\ }\textbf {\bibinfo
  {volume} {124}},\ \bibinfo {pages} {1866--1878} (\bibinfo {year}
  {1961})}\BibitemShut {NoStop}%
\bibitem [{\citenamefont {Wickenhauser}\ \emph {et~al.}(2005)\citenamefont
  {Wickenhauser}, \citenamefont {Burgd\"orfer}, \citenamefont {Krausz},\ and\
  \citenamefont {Drescher}}]{Wickenhauser05}%
  \BibitemOpen
  \bibfield  {author} {\bibinfo {author} {\bibfnamefont {Marlene}\ \bibnamefont
  {Wickenhauser}}, \bibinfo {author} {\bibfnamefont {Joachim}\ \bibnamefont
  {Burgd\"orfer}}, \bibinfo {author} {\bibfnamefont {Ferenc}\ \bibnamefont
  {Krausz}}, \ and\ \bibinfo {author} {\bibfnamefont {Markus}\ \bibnamefont
  {Drescher}},\ }\bibfield  {title} {\enquote {\bibinfo {title} {Time resolved
  fano resonances},}\ }\href {\doibase 10.1103/PhysRevLett.94.023002}
  {\bibfield  {journal} {\bibinfo  {journal} {Phys. Rev. Lett.}\ }\textbf
  {\bibinfo {volume} {94}},\ \bibinfo {pages} {023002} (\bibinfo {year}
  {2005})}\BibitemShut {NoStop}%
\bibitem [{\citenamefont {Kramida}\ \emph {et~al.}(2018)\citenamefont
  {Kramida}, \citenamefont {Ralchenko}, \citenamefont {Reader},\ and\
  \citenamefont {(2018)}}]{NIST2018}%
  \BibitemOpen
  \bibfield  {author} {\bibinfo {author} {\bibfnamefont {A.}~\bibnamefont
  {Kramida}}, \bibinfo {author} {\bibfnamefont {Yu.}\ \bibnamefont
  {Ralchenko}}, \bibinfo {author} {\bibfnamefont {J.}~\bibnamefont {Reader}}, \
  and\ \bibinfo {author} {\bibfnamefont {NIST ASD~Team}\ \bibnamefont
  {(2018)}},\ }\href@noop {} {\enquote {\bibinfo {title} {Nist atomic spectra
  database (version 5.6.1)},}\ } (\bibinfo {year} {2018}),\ \bibinfo {note}
  {online}\BibitemShut {NoStop}%
\bibitem [{\citenamefont {Ghosh}\ and\ \citenamefont {Vaval}(2014)}]{Ghosh14b}%
  \BibitemOpen
  \bibfield  {author} {\bibinfo {author} {\bibfnamefont {A.}~\bibnamefont
  {Ghosh}}\ and\ \bibinfo {author} {\bibfnamefont {N.}~\bibnamefont {Vaval}},\
  }\bibfield  {title} {\enquote {\bibinfo {title} {{Geometry-dependent lifetime
  of Interatomic coulombic decay using equation-of-motion coupled cluster
  method}},}\ }\href {\doibase http://dx.doi.org/10.1063/1.4903827} {\bibfield
  {journal} {\bibinfo  {journal} {J. Chem. Phys.}\ }\textbf {\bibinfo {volume}
  {141}},\ \bibinfo {pages} {234108} (\bibinfo {year} {2014})}\BibitemShut
  {NoStop}%
\bibitem [{\citenamefont {Bondi}(1964)}]{Bondi64}%
  \BibitemOpen
  \bibfield  {author} {\bibinfo {author} {\bibfnamefont {A.}~\bibnamefont
  {Bondi}},\ }\bibfield  {title} {\enquote {\bibinfo {title} {van der waals
  volumes and radii},}\ }\href@noop {} {\bibfield  {journal} {\bibinfo
  {journal} {J. Phys. Chem.}\ }\textbf {\bibinfo {volume} {68}},\ \bibinfo
  {pages} {441} (\bibinfo {year} {1964})}\BibitemShut {NoStop}%
\bibitem [{ELD()}]{ELDEST_v2}%
  \BibitemOpen
  \href {\doibase 10.5281/zenodo.3700062} {}\bibinfo {note} {ELDEST version
  2.0.3, Programme for the time-resolved investigation of electronic decay
  processes (2020), written by E. Fasshauer,
  DOI:10.5281/zenodo.3700062}\BibitemShut {NoStop}%
\bibitem [{\citenamefont {Jones}\ \emph {et~al.}(2001--)\citenamefont {Jones},
  \citenamefont {Oliphant}, \citenamefont {Peterson} \emph {et~al.}}]{SciPy}%
  \BibitemOpen
  \bibfield  {author} {\bibinfo {author} {\bibfnamefont {E.}~\bibnamefont
  {Jones}}, \bibinfo {author} {\bibfnamefont {T.}~\bibnamefont {Oliphant}},
  \bibinfo {author} {\bibfnamefont {P.}~\bibnamefont {Peterson}},  \emph
  {et~al.},\ }\href {http://www.scipy.org/} {\enquote {\bibinfo {title}
  {{SciPy}: Open source scientific tools for {Python}},}\ } (\bibinfo {year}
  {2001--})\BibitemShut {NoStop}%
\bibitem [{\citenamefont {Oliphant}(2007)}]{Numpy1}%
  \BibitemOpen
  \bibfield  {author} {\bibinfo {author} {\bibfnamefont {Travis~E.}\
  \bibnamefont {Oliphant}},\ }\bibfield  {title} {\enquote {\bibinfo {title}
  {Python for scientific computing},}\ }\href {\doibase 10.1109/MCSE.2007.58}
  {\bibfield  {journal} {\bibinfo  {journal} {Computing in Science \&
  Engineering}\ }\textbf {\bibinfo {volume} {9}},\ \bibinfo {pages} {10}
  (\bibinfo {year} {2007})}\BibitemShut {NoStop}%
\bibitem [{\citenamefont {Millman}\ and\ \citenamefont
  {Aivazis}(2011)}]{Numpy2}%
  \BibitemOpen
  \bibfield  {author} {\bibinfo {author} {\bibfnamefont {K.~Jarrod}\
  \bibnamefont {Millman}}\ and\ \bibinfo {author} {\bibfnamefont {Michael}\
  \bibnamefont {Aivazis}},\ }\bibfield  {title} {\enquote {\bibinfo {title}
  {Python for scientists and engineers},}\ }\href {\doibase
  10.1109/MCSE.2011.36} {\bibfield  {journal} {\bibinfo  {journal} {Computing
  in Science \& Engineering}\ }\textbf {\bibinfo {volume} {13}},\ \bibinfo
  {pages} {9} (\bibinfo {year} {2011})}\BibitemShut {NoStop}%
\bibitem [{\citenamefont {Fasshauer}(2019)}]{rapid_input}%
  \BibitemOpen
  \bibfield  {author} {\bibinfo {author} {\bibfnamefont {E.}~\bibnamefont
  {Fasshauer}},\ }\href@noop {} {} (\bibinfo {year} {2019}),\ \bibinfo {note}
  {input file, DOI:10.6084/m9.figshare.7808348.v1}\BibitemShut {NoStop}%
\bibitem [{vid()}]{video}%
  \BibitemOpen
  \href@noop {} {}\bibinfo {note} {Supplementary video of the time-resolved
  spectrum sRICD and ICD spectrum, where the second pulse quenches the sRICD
  process after 100 fs.}\BibitemShut {Stop}%
\bibitem [{\citenamefont {Wickenhauser}(2006)}]{Wickenhauser_thesis}%
  \BibitemOpen
  \bibfield  {author} {\bibinfo {author} {\bibfnamefont {M.}~\bibnamefont
  {Wickenhauser}},\ }\emph {\bibinfo {title} {Ionization Dynamics of Atoms in
  Femto- and Attosecond Pulses}},\ \href@noop {} {Ph.D. thesis},\ \bibinfo
  {school} {Vienna University of Technology} (\bibinfo {year}
  {2006})\BibitemShut {NoStop}%
\bibitem [{\citenamefont {Busto}\ \emph {et~al.}(2018)\citenamefont {Busto},
  \citenamefont {Barreau}, \citenamefont {Isinger}, \citenamefont {Turconi},
  \citenamefont {Alexandridi}, \citenamefont {Harth}, \citenamefont {Zhong},
  \citenamefont {Squibb}, \citenamefont {Kroon}, \citenamefont {Plogmaker},
  \citenamefont {Miranda}, \citenamefont {Jimenez-Galan}, \citenamefont
  {Argenti}, \citenamefont {Arnold}, \citenamefont {Feifel}, \citenamefont
  {Martin}, \citenamefont {Gisselbrecht}, \citenamefont {L'Huillier},\ and\
  \citenamefont {Salieres}}]{Busto18}%
  \BibitemOpen
  \bibfield  {author} {\bibinfo {author} {\bibfnamefont {D.}~\bibnamefont
  {Busto}}, \bibinfo {author} {\bibfnamefont {L.}~\bibnamefont {Barreau}},
  \bibinfo {author} {\bibfnamefont {M.}~\bibnamefont {Isinger}}, \bibinfo
  {author} {\bibfnamefont {M.}~\bibnamefont {Turconi}}, \bibinfo {author}
  {\bibfnamefont {C.}~\bibnamefont {Alexandridi}}, \bibinfo {author}
  {\bibfnamefont {A.}~\bibnamefont {Harth}}, \bibinfo {author} {\bibfnamefont
  {S.}~\bibnamefont {Zhong}}, \bibinfo {author} {\bibfnamefont {R.~J.}\
  \bibnamefont {Squibb}}, \bibinfo {author} {\bibfnamefont {D.}~\bibnamefont
  {Kroon}}, \bibinfo {author} {\bibfnamefont {S.}~\bibnamefont {Plogmaker}},
  \bibinfo {author} {\bibfnamefont {M.}~\bibnamefont {Miranda}}, \bibinfo
  {author} {\bibfnamefont {A.}~\bibnamefont {Jimenez-Galan}}, \bibinfo {author}
  {\bibfnamefont {L.}~\bibnamefont {Argenti}}, \bibinfo {author} {\bibfnamefont
  {C.~L.}\ \bibnamefont {Arnold}}, \bibinfo {author} {\bibfnamefont
  {R.}~\bibnamefont {Feifel}}, \bibinfo {author} {\bibfnamefont
  {F.}~\bibnamefont {Martin}}, \bibinfo {author} {\bibfnamefont
  {M.}~\bibnamefont {Gisselbrecht}}, \bibinfo {author} {\bibfnamefont
  {A.}~\bibnamefont {L'Huillier}}, \ and\ \bibinfo {author} {\bibfnamefont
  {P.}~\bibnamefont {Salieres}},\ }\bibfield  {title} {\enquote {\bibinfo
  {title} {Time–frequency representation of autoionization dynamics in
  helium},}\ }\href@noop {} {\bibfield  {journal} {\bibinfo  {journal} {J.
  Phys. B: At. Mol. Opt. Phys.}\ }\textbf {\bibinfo {volume} {51}},\ \bibinfo
  {pages} {044002} (\bibinfo {year} {2018})}\BibitemShut {NoStop}%
\end{thebibliography}%

\end{document}